\documentclass[10pt,fleqn,a4paper,twoside]{article}

\usepackage[sort&compress,numbers]{natbib}
\usepackage[margin=1in]{geometry}
\usepackage{amsmath,amssymb,amsfonts}
\usepackage[version=4]{mhchem}
\usepackage{booktabs}
\usepackage{authblk}
\usepackage[english]{babel}
\usepackage{graphicx}
\usepackage{siunitx}
\usepackage{longtable,tabularx}
\usepackage[labelformat=simple]{subcaption}

\newcommand{\Reyn}{\operatorname{\mathit{Re}}}
\newcommand{\ReynCell}{\operatorname{\mathit{Re}}_{cell}}
\newcommand{\Mach}{\operatorname{\mathit{Ma}}}

\newcommand{\Knud}{\operatorname{\mathit{Kn}}}
\newcommand{\Lewi}{\operatorname{\mathit{Le}}}

\begin{document}

\title{
	\MakeUppercase{Influence of Park's Two-Temperature Model Control Temperature on the Flow Properties in Hypersonic Reentry Conditions}
}

\author[1]{{Gibson De Marchi Poltronieri}\thanks{Corresponding author: gibsondemarchi@gmail.com}}
\affil[1]{\small{Instituto Tecnológico de Aeronáutica, São José dos Campos, SP, 12228--900\@, Brazil}}

\author[1]{{Farney C. Moreira} \thanks{farney.coutinho@gmail.com}}

\author[2]{{João Luiz F. Azevedo}\thanks{joaoluiz.azevedo@gmail.com}}
\affil[2]{\small{Instituto de Aeronáutica e Espaço, São José dos Campos, SP, 12228--904\@, Brazil}}

\date{}

\maketitle

\begin{abstract}
	Numerical simulations of reactive hypersonic flows under thermochemical non-equilibrium conditions are presented for the {FIRE II} and Mars Pathfinder capsules. An 11-species chemical model is employed to simulate Earth's atmosphere, while an 8-species chemical model simulates Mars' atmosphere. The current formulation uses Park's two-temperature model to account for the non-equilibrium phenomena. The present work analyzes the impact of different sets of weight factors used in Park's model to calculate the control temperature. The code used to simulate the hypersonic flow addressed in this work solves the Navier-Stokes equations for reacting gas flows. The findings are depicted in terms of the Mach number, temperature modes, and mass fraction distributions along the stagnation streamline in a region closer to the shock wave. The study also includes results regarding the stagnation point convective heat flux. The results presented are encouraging and show that the weight factors significantly impact the {FIRE II} test cases while having little impact on the Mars Pathfinder flows. In all cases, it is possible to observe some effect of the weight factor selection on property distributions. In summary, the weight factors influence the flow behavior with varying intensities depending on the flow conditions.

		\textit{CFD},
		\textit{Thermochemical non-equilibrium},
		\textit{Heat transfer},
		\textit{Chemical reactions}
\end{abstract}

\section{INTRODUCTION\label{sec:introduction}}

Atmospheric entry/reentry is one of the most challenging topics for engineers in space exploration. In recent years, public agencies and private corporations have reignited the interest in investing in space exploration. The National Aeronautics and Space Administration contracted with private organizations for the research, design, and testing of spacecraft for future space missions, for example. The entry/reentry spacecraft needs to perform maneuvers in high-speed conditions, typically at speeds of the order of \SI[per-mode=symbol]{10}{\km\per\s}\@. The atmospheric entry/reentry procedure aims to use the atmosphere itself to decelerate the vehicles to suitable lower speeds for a safe landing. During this procedure, the spacecraft must withstand the severe flow conditions imposed by the high-speed flow and the varying density regimes imposed by the different atmospheric layers. The physicochemical complexity of the flow increases significantly due to high temperature, leading to the appearance of non-equilibrium phenomena. Chemical reactions, such as dissociation, recombination, and ionization, occurring within the flow change the gas mixture composition and, consequently, the physicochemical properties of the flow. Therefore, the design of hypersonic vehicles, including the Thermal Protection System (TPS), must account for those many complex phenomena accordingly.

The present work aims to evaluate the impact of the weight factors of the control temperature from Park's two-temperature model on the flow behavior. The authors are motivated by the fact that the literature has little in-depth work about those weight factors. The present work assesses the impact of the weight factors in two different configurations, addressing different gas mixture compositions. The first configuration is the {FIRE II} reentry capsule reentering Earth's atmosphere. The second case considers the Mars Pathfinder capsule entering Mars' atmosphere. The present work obtained the results through the {LeMANS} parallel code~\cite{lc_scalabrin_phd_2007}, which solves the Navier-Stokes equations and accounts for both thermodynamic and chemical non-equilibrium. The work compares the present numerical results with numerical and experimental data available in the literature~\cite{cornette_1966,hollis_1996,fc_moreira_phd_2020,niu_et_al_2018}.

To assess the impact of the weight factors, the present work uses a set of weight factors that encompass the values regarded as good choices by the literature~\cite{c_park_2010,niu_et_al_2018,moreira_wolf_azevedo_scitech_2023,moreira_wolf_azevedo_jhmt_2021}. This study presents the numerical results in terms of the Mach number, temperature modes, and mass fraction distributions along the stagnation streamline, focusing on the non-equilibrium phenomena occurring near the shock wave. Moreover, the present analysis includes data for the stagnation point convective heat flux. Therefore, the present work aims to provide a broader understanding of the influence of the weight factors of the control temperature of Park's two-temperature model on the behavior of reactive hypersonic flows.

\section{THEORETICAL FORMULATION\label{sec:theoretical_formulation}}

\subsection{General Considerations}

The freestream density is a function of the altitude in the atmosphere. Therefore, the entry/reentry vehicle experiences different flow regimes based on the freestream density throughout the trajectory. The Knudsen number is a dimensionless parameter often employed to differentiate flow regimes in continuum, transition, and free-molecule. This classification is of utmost importance to define the physicomathematical formulation most suited to model the flow~\cite{boyd_schwartzentruber_2017,agrawal_2020}. The Knudsen number is defined as
\begin{equation}
    \Knud = \frac{\lambda}{L}
    \qquad \text{,}
\end{equation}
where \( \lambda \) is the mean free path, which means the average distance traveled by a molecule, or atom, between successive collisions, and the \( L \) is the characteristic length, which is a representative measure of the dimension of the fluid-immersed body~\cite{ga_bird_1994}. The Knudsen number can also be rewritten as a function of the Reynolds number, Mach number, and ratio of specific heats as
\begin{equation}
    \Knud = \frac{\Mach}{\Reyn} \sqrt{\frac{\gamma \pi}{2}}
    \qquad \text{,}
\end{equation}
by assuming the Hard-Sphere (H-S) model for particle collision cross section~\cite{boyd_schwartzentruber_2017}. In the above equation, \( \Reyn \) is the Reynolds number, \( \Mach \) is the Mach number, and \( \gamma \) is the ratio of specific heats.

As stated, the Knudsen number is used to differentiate flow regimes. Therefore, sufficiently low Knudsen number values indicate that the flow is in a continuum regime, while very high Knudsen number values imply the free-molecule regime. The transition regime is between the continuum and the free-molecule regime. The literature often proposes varying Knudsen number values to differentiate these flow regimes. Reference~\cite{ga_bird_2013} defines the transition regime as the region between the slip and free-molecule flow regimes. Reference~\cite{boyd_schwartzentruber_2017} states that the transition regime lies between \( \num{e-02} \leq \Knud < 10 \)\@, whereas Ref.~\cite{anderson_2006} defines the transition regime within \( \num{3e-02} \leq \Knud < 1 \)\@. Reference~\cite{agrawal_2020} uses \( \num{e-01} \leq \Knud < 10 \) to define the transition regime. These differences indicate that the limiting Knudsen number values that define the transition flow regime remain unsettled in the literature.

The well-known Navier-Stokes equations can accurately model the continuum regime, including high-enthalpy flows. Additionally, the Navier-Stokes equations are also suited to model the slip flow regime, which is a subset of the continuum flow regime that is closer to the transition regime. For Knudsen number values within the transition and free-molecule regimes, the Navier-Stokes equations lose their capabilities of representing the physics of the flow accurately. In these cases, higher-order continuum transport formulations may be used to extend the viability of the continuum formulations for higher Knudsen number values. Reference~\cite{agrawal_2020} presents the Burnett, super-Burnett, Grad, and Onsager-Burnett equations as some examples of high-order continuum formulations.

Alternatively, the transition and free-molecule flow regimes are modeled by discrete formulations. The Boltzmann equation and its variants are widely employed to describe the transition regime. The collisionless Boltzmann equation, for example, is appropriate for solving the free-molecule regime, where energy changes may occur without collisions. Moreover, the discrete approach is not limited to high Knudsen number values and can accurately represent the physics of the flow in the continuum flow regime. The Direct Simulation Monte Carlo (DSMC) is a collision-based method to solve flow regimes in the transition and free-molecule regimes~\cite{ga_bird_1994,ga_bird_2013}.

The hypersonic flows addressed in the present study have characteristic freestream Knudsen number values sufficiently low~\cite{moreira_wolf_azevedo_jhmt_2021}. Therefore, the flows studied in this work are within the continuum regime, allowing the use of the Navier-Stokes equations. Furthermore, the present work uses Park's two-temperature model to account for thermodynamic non-equilibrium and weak ionization effects~\cite{c_park_1989}. Park's model couples the temperatures of the translational and rotational energy modes into a single translational-rotational temperature mode, called \( T_{tr} \)\@. The second temperature of Park's model is the vibrational-electronic temperature mode, \( T_{ve} \)\@, associated with the coupling of the vibrational and electronic energy modes, along with the electron energy~\cite{lc_scalabrin_phd_2007,martin_scalabrin_boyd_2012}.

\subsection{Conservation Equations and Related Models}

The present analysis employs the Navier-Stokes equations for chemically reacting gas flows. The system of conservation equations contains two source terms representing the chemical reactions, including non-equilibrium phenomena, \( S_{c,v} \)\@, and axisymmetric flow configuration, \( S_{axi} \)\@. Therefore, the Navier-Stokes equations can be written in a multi-dimensional form using the index notation as
\begin{equation}
    \frac{\partial Q}{\partial t}
    + \frac{\partial (F_{j} - F_{v_{j}})}{\partial x_{j}}
    = S_{c,v} + S_{axi}
    \qquad \text{,}
    \label{eqn:navier_stokes}
\end{equation}
where \( Q \) is the vector of conserved variables
\begin{equation}
    Q = 
    \begin{Bmatrix}
        \rho_{1} & \dots & \rho_{ns} & \rho u_{i} & E & E_{ve}
    \end{Bmatrix}^{\mathsf{T}}
    \qquad \text{.}
    \label{eqn:conservated_variables}
\end{equation}
In the index notation, a free index represents a vectorial equation, and repeated indices indicate a summation. Therefore, in the definition of \( Q \)\@, \( \rho_{1},\ \dots,\ \rho_{ns} \) are the density of the \( ns \) chemical species, \( \rho \) is the gas mixture density, \( u_{i} \) are the velocity components, \( E \) is the total energy per unit volume of the gas mixture, and \( E_{ve} \) is the vibrational-electronic energy per unit volume of the gas mixture.

The inviscid and viscous flux terms, \( F_{j} \) and \( F_{v_{j}} \)\@, respectively, are defined as 
\begin{equation}
    F_{j} = \begin{Bmatrix}
        \rho_{1} u_{j} \\
        \vdots \\
        \rho_{ns} u_{j} \\
        \rho u_{i} u_{j} + p \delta_{ij} \\
        \left(E + p\right) u_{j} \\
        E_{ve} u_{j}
    \end{Bmatrix}
    \qquad \qquad
    \text{and}
    \qquad \qquad
    F_{v_{j}} = \begin{Bmatrix}
        - J_{1,j} \\
        \vdots \\
        - J_{ns,j} \\
        \tau_{ij} \\
        \tau_{ij} - \left(q_{tr,j} + q_{ve,j}\right) - \sum \left(J_{s,j} h_{s}\right) \\
        - q_{ve,j} - \sum \left(J_{s,j} e_{ve,s}\right)
    \end{Bmatrix}
    \qquad \text{.}
\end{equation}
In the definition of \( F_{j} \)\@, \( p \) is the gas mixture pressure and \( \delta_{ij} \) is the Kronecker's delta function. In the definition of \( F_{v_{j}} \)\@, \( J_{1,j},\ \dots,\ J_{ns,j} \) are the diffusion flux of the \( ns \) chemical species in the \( j \)-th direction, \( \tau_{ij }\) is the viscous stress tensor components, \( q_{tr,j} \) is the translational-rotational heat flux in the \( j \)-th direction, \( q_{ve,j} \) is the vibrational-electronic heat flux in the \( j \)-th direction, \( h_{s} \) is the enthalpy of the \( s \)-th chemical species, and \( e_{ve,s} \) is the specific vibrational-electronic energy of the \( s \)-th chemical species.

The diffusion flux, \( J_{ns,j} \)\@, is given by Fick's Law as
\begin{equation}
    J_{s \neq e,j} = \rho D_{s} \frac{\partial Y_{s}}{\partial x_{j}}
    \qquad \text{,}
\end{equation}
where \( D_{s} \) is the diffusion coefficient and \( Y_{s} \) is the mass fraction of the \( s \)-th chemical species, except the electron. The diffusion flux for the electrons is calculated by considering ambipolar diffusion to guarantee charge neutrality~\cite{lc_scalabrin_phd_2007}. Considering that Dalton's Law of partial pressure is valid and that each chemical species behaves as an ideal gas, the gas mixture pressure, \( p \)\@, can be written as 
\begin{equation}
    p = 
    \sum\limits_{s \neq e} \left(\frac{\rho_{s} R_{u}}{M_{s}} T_{tr}\right)
    + \frac{\rho_{e} R_{u}}{M_{e}} T_{ve}
    \qquad \text{,}
\end{equation}
where \( R_{u} \) is the universal gas constant~\cite{gillespie_1930}. The viscous stress tensor, for a Newtonian fluid, is given by
\begin{equation}
    \tau_{ij} 
    = \mu \left(
        \frac{\partial u_{i}}{\partial x_{j}}
        + \frac{\partial u_{j}}{\partial x_{i}}
    \right)
    - \left(\frac{2}{3}\mu - \beta\right) 
    \frac{\partial u_{k}}{\partial x_{k}} \delta_{ij}
    \qquad \text{,}
\end{equation}
where \( \mu \) is the mixture coefficient of viscosity and \( \beta \) is the bulk viscosity. The bulk viscosity arises from the momentum exchange between colliding molecules and their internal degrees of freedom, contributing to the dilatational term that appears in the normal stress. Therefore, the bulk viscosity could directly impact the calculations in the present formulation, especially in carbon dioxide flows. The literature presents some methods to calculate the bulk viscosity under different temperature ranges~\cite{cramer_2012,jaeger_matar_muller_2018,sharma_kumar_2023}. However, those models are usually for low-temperature values. Therefore, the present formulation assumes the Stokes' hypothesis, \( \beta = 0 \)\@. This assumption aims to avoid inaccuracies during the bulk viscosity calculations, which could consequently degrade the numerical results obtained. The present formulation employs Fourier's Law to model the heat fluxes as
\begin{equation}
    q_{tr,j} = - \kappa_{tr} \frac{\partial T_{tr}}{\partial x_{j}}
    \qquad \qquad \text{and} \qquad \qquad
    q_{ve,j} = - \kappa_{ve} \frac{\partial T_{ve}}{\partial x_{j}}
    \qquad \text{,}
\end{equation}
where \( \kappa_{tr} \) and \( \kappa_{ve} \) are the thermal conductivity coefficients of the gas mixture associated with the translational-rotational and vibrational-electronic temperature modes, respectively.

This work employs two approaches to calculate the transport properties~\cite{lc_scalabrin_phd_2007}. The first approach uses Ref.~\cite{wilke_1950} semi-empirical mixing rule, Ref.~\cite{blottner_johnson_ellis_1971} curve fits, and Eucken's relation~\cite{vincenti_kruger_1982,anderson_2006}. The second approach is the one proposed in Ref.~\cite{gupta_yos_thompson_lee_1990}. The first approach is suited for low velocity flows with maximum temperatures around \SI[per-mode=symbol]{10000}{\kelvin} and is not designed for ionized flows. The second model is suitable for high-speed, around \SI[per-mode=symbol]{10}{\km\per\s}, and weakly ionized flows. Considering the first approach, the gas mixture viscosity and thermal conductivity coefficients are given by~\cite{wilke_1950}
\begin{equation}
    \mu = \sum\limits_{s}^{ns} \frac{X_{s} \mu_{s}}{\phi_{s}}
    \qquad \qquad \text{and} \qquad \qquad
    \kappa = \sum\limits_{s}^{ns} \frac{X_{s} \kappa_{s}}{\phi_{s}}
    \qquad \text{,}
\end{equation}
respectively. In the above equation, the \( \phi_{s} \) term is given by
\begin{equation}
    \phi_{s} = 
    \sum\limits_{r}^{nr} X_{r}
    \left[
        1 + \sqrt{\frac{\mu_{s}}{\mu_{r}}}
        \left(\frac{M_{r}}{M_{s}}\right)^{1/4}
    \right]^{2}
    \left[
        \sqrt{8 \left(1 + \frac{M_{s}}{M_{r}}\right)}
    \right]^{-1}
    \qquad \text{,}
\end{equation}
where \( X \) is the molar fraction and \( M \) is the molar weight of the \( s \)-th and \( r \)-th chemical species. The dynamic viscosity coefficients, \( \mu_{s} \)\@, are calculated by Blottner's curve fits for each chemical species as
\begin{equation}
    \mu_{s} = 0.1 \exp\left\{
        \left[A_{s} \ln(T) + B_{s}\right]
        \ln(T) + C_{s}
    \right\}
    \qquad \text{,}
\end{equation}
where \( A_{s} \)\@, \( B_{s} \)\@, and \( C_{s} \) are constants~\cite{blottner_johnson_ellis_1971}. The thermal conductivity coefficients, \( \kappa_{tr} \) and \( \kappa_{ve} \)\@, are calculated, also for each chemical species, by Eucken's relation as
\begin{equation}
    \kappa_{tr,s} = \mu_{s} \left( \frac{5}{2} C_{v_{t,s}} + C_{v_{r,s}} \right) 
    \qquad \qquad \text{and} \qquad \qquad
    \kappa_{ve,s} = \mu_{s} C_{v_{ve,s}}
    \qquad \text{,}
\end{equation}
where \( C_{v} \) is the specific heat at constant volume for different internal energy modes of the chemical species~\cite{vincenti_kruger_1982,anderson_2006}. The diffusion coefficient for each chemical species, \( D_{s} \)\@, is calculated simply by a single binary coefficient \( D \)\@, which ensures that the sum of the diffusion fluxes is zero. Thus, given by
\begin{equation}
    D = \frac{\kappa_{tr}}{\rho C_{p_{tr}}} \Lewi
    \qquad \text{,}
\end{equation}
where \( C_{p_{tr}} \) is the gas mixture specific heat at constant pressure associated with the translational-rotational temperature mode and \( \Lewi \) is the Lewis number, a dimensionless number that represents the ratio between thermal diffusivity and mass diffusivity. As stated, this approach is not accurate for velocities above \SI[per-mode=symbol]{10}{\km\per\s}~\cite{lc_scalabrin_phd_2007}. For the sake of brevity, the present work will not describe the second approach to calculate the transport properties. References~\cite{lc_scalabrin_phd_2007,gupta_yos_thompson_lee_1990} provide more details about the second approach.

In Eq.~\eqref{eqn:navier_stokes}, the source term, \( S_{axi} \)\@, represents the additional surface stress that appears by an axisymmetric formulation. The contribution is only to the y-momentum equation to counterbalance the pressure and viscous forces acting on the side of the surface of the control volume. Therefore, the \( S_{axi} \) term is given by
\begin{equation}
    S_{axi} = 
    \begin{Bmatrix}
        0 & \dots & 0 & 0 & \left[
            -p + 2\mu \left(
                \dfrac{u_{2}}{\bar{x}_{2}}
                - \dfrac{1}{3} \dfrac{\partial u_{k}}{\partial x_{k}}
            \right)
        \right] \dfrac{\delta_{i2}}{\bar{x}_{2}} & 0 & 0
    \end{Bmatrix}^{\mathsf{T}}
    \qquad \text{,}
\end{equation}
where 
\begin{equation}
    \frac{\partial u_{k}}{\partial x_{k}}
    = \frac{\partial u_{1}}{\partial x_{1}}
    + \frac{\partial u_{2}}{\partial x_{2}}
    + \frac{u_{2}}{\bar{x}_{2}}
    \qquad \text{.}
\end{equation}
In the above formulation, the \( \bar{x}_{2} \) is the radial coordinate measure from the axis of symmetry to the cell centroid and \( x_{1} \) and \( x_{2} \) are the axial and radial directions, respectively.

The source term \( S_{c,v} \) in Eq.~\eqref{eqn:navier_stokes} is associated with the rate of production or consumption of mass of the chemical species in the reacting flow and the energy related to the vibrational energy mode. Therefore, the \( S_{c,v} \) source term is written as
\begin{equation}
    S_{c,v} = 
    \begin{Bmatrix}
        \dot{w}_{1} & \dots & \dot{w}_{ns} & 0 & 0 & 0 & 0 & \dot{w}_{v}
    \end{Bmatrix}^{\mathsf{T}}
    \qquad \text{,}
    \label{eqn:chem_source_term}
\end{equation}
where \( \dot{w}_{1},\ \dots,\ \dot{w}_{ns} \) are the rate of mass production of the \( ns \) chemical species and \( \dot{w}_{v} \) is the vibrational energy source term.

\subsection{Chemical Species and Model of Chemical Reactions}

The flow gas mixture composition changes due to chemical reactions occurring in high-enthalpy hypersonic flows. These changes impact the physicochemical properties of the flow. Therefore, chemical models have been developed to represent those high-enthalpy flow phenomena according to the flow physicochemical complexity~\cite{gnoffo_gupta_shinn_1989}. The present work uses an 11-species chemical model to represent Earth's atmosphere and an 8-species chemical model to represent the Mars atmosphere. Table~\ref{tab:chemical_models} presents the chemical species that compose the chemical models used for the Earth and Mars atmospheres.

\begin{table}[hbt!]
    \caption{\label{tab:chemical_models} Chemical species of Earth and Mars atmosphere models.}
    \centering
    \begin{tabular}{lc}
    \hline
        Chemical Model & Chemical Species \\\hline
        11-species (Earth)
            & \ce{N2}\@, \ce{O2}\@, \ce{NO}\@, \ce{N}\@, \ce{O}\@, \ce{NO+}\@, \ce{N2+}\@, \ce{O2+}\@, \ce{N+}\@, \ce{O+}\@, and \ce{e-} \\
        8-species (Mars)
            & \ce{CO2}\@, \ce{CO}\@, \ce{N2}\@, \ce{O2}\@, \ce{NO}\@, \ce{N}\@, \ce{O}\@, and \ce{C} \\
    \hline
    \end{tabular}
\end{table}

The present work considers a finite-rate chemistry model for the reacting gas mixture of the hypersonic entry/reentry flows studied~\cite{lc_scalabrin_phd_2007}. The chemical reactions are generically written as
\begin{equation}
    \sum \alpha_{s,r} \left[ S \right]
    \ce{<=>}
    \sum \beta_{s,r} \left[ S \right]
    \qquad \text{,}
\end{equation}
where \( \left[ S \right] \) represents the chemical species listed in Table~\ref{tab:chemical_models} and \( \alpha_{s,r} \) and \( \beta_{s,r} \) represents the stoichiometric coefficients of the \(s\)-th chemical species that balance the \(r\)-th chemical reaction considered in the chemistry model. The detailed list of chemical reactions considered in the chemical models for the Earth and Mars atmospheres are found in Refs.~\cite{lc_scalabrin_phd_2007,fc_moreira_phd_2020}. The present work standardizes the forward reactions as endothermic reactions and the backward chemical reactions as exothermic reactions. The endothermic chemical reactions absorb energy from the surroundings while exothermic chemical reactions release energy to the surroundings~\cite{atkins_paula_2006}. The rate of mass production, or consumption, of the \(s\)-th chemical species, in Eq.~\eqref{eqn:chem_source_term}\@, is given by~\cite{lc_scalabrin_phd_2007}
\begin{equation}
    \dot{w}_{s} = 
    M_{s} \sum\limits_{r}^{nr} 
    \left(\beta_{s,r} - \alpha_{s,r}\right) 
    \left[
    k_{f,r} \prod\limits_{s}^{ns} 
        \left(
            \frac{\rho_{s}}{M_{s}}
        \right)^{\alpha_{s,r}}
        -
    k_{b,r} \prod\limits_{s}^{ns} 
        \left(
            \frac{\rho_{s}}{M_{s}}
        \right)^{\beta_{s,r}}
    \right]
    \qquad \text{,}
\end{equation}
where \( nr \) represents the number of chemical reactions that the \( s \)-th chemical species participates and \( k_{f,r} \) and \( k_{b,r} \) are the forward and backward chemical reaction rates of the \( r \)-th chemical reaction, respectively.

The thermodynamic and chemical non-equilibrium phenomena occur when the characteristic time scale of the flow is in the same order of magnitude as the relaxation time of the system to the equilibrium state. The thermodynamic and chemical non-equilibrium in the flow affects the forward and backward chemical reaction rates. The present work uses Park's two-temperature model, Ref.~\cite{c_park_1989}\@, to account for the non-equilibrium phenomena in the hypersonic entry/reentry flows studied. In Park's model, the translational and rotational temperature modes are coupled into the translational-rotational temperature mode, \( T_{tr} \)\@, and the vibrational, electronic, and electron translational temperature modes are coupled into the vibrational-electronic temperature mode, \( T_{ve} \)\@. Moreover, Park's model defines the control temperature, \( T_{c} \)\@, given as a combination of \( T_{tr} \) and \( T_{ve} \) by an geometric weighted mean as
\begin{equation}
    T_{c} = T_{tr}^{a} T_{ve}^{b}
    \qquad \text{,}
\end{equation}
where \( a \) and \( b = 1 - a \) are weight factors that control the energy transfer between dissociation and ionization reactions. It must be noted that the control temperature, \( T_{c} \)\@, can assume the value of \( T_{tr} \)\@, \( T_{ve} \)\@, or \( T_{tr}^{a} T_{ve}^{b} \) depending on the of chemical reaction~\cite{lc_scalabrin_phd_2007,niu_et_al_2018}. This formulation accounts for the fact that vibrationally excited molecules are more likely to dissociate~\cite{c_park_1989}. The typical values for \( a \) and \( b \) indicated by the literature are \( a = b = 0.5 \) or \( a = 0.7,\ b = 0.3 \)~\cite{lc_scalabrin_phd_2007,c_park_2010}.

The forward reaction rate of the \(r\)-th chemical reaction, \( k_{f,r} \)\@, is a function of the control temperature, \( T_{c} \)\@, and is calculated by Arrhenius curve fits of the form
\begin{equation}
    k_{f,r} (T_{c})
    = C_{f,r} T_{c}^{\eta_{r}} \exp\left(- \frac{\theta_{r}}{T_{c}}\right)
    \qquad \text{,}
\end{equation}
where \( C_{f,r} \) is the pre-exponential factor, \( \eta_{r} \) is the temperature dependence, and \( \theta_{r} \) is the activation energy. The forward chemical kinetic rate coefficients are constants proposed in Ref.~\cite{c_park_1990}. The backward reaction rate, \( k_{b,r} \)\@, is given by
\begin{equation}
    k_{b,r} (T_{c}) = \frac{k_{f,r} (T_{c})}{k_{eq,r} (T_{c})}
    \qquad \text{,}
\end{equation}
where \( k_{eq,r} \) is the equilibrium constant of the \( r \)-th chemical reaction~\cite{atkins_paula_2006}. The equilibrium constant, \( k_{eq} \)\@, is calculated by curve fits as follows
\begin{equation}
    k_{eq,r} (T_{c})
    =
    \exp \left[
        A_{1} \left(\frac{T_{c}}{10^{4}}\right)
        + A_{2}
        + A_{3} \ln \left(\frac{10^{4}}{T_{c}}\right)
        + A_{4} \left(\frac{10^{4}}{T_{c}}\right)
        + A_{5} \left(\frac{10^{4}}{T_{c}}\right)^{2}
    \right]
    \qquad \text{,}
\end{equation}
where the coefficients \( A_{1} \)\@, \( A_{2} \)\@, \( A_{3} \)\@, \( A_{4} \)\@, and \( A_{5} \) are functions of the local number density of the flow within the range of data tabulated in Ref.~\cite{c_park_1989}. In cases where the number density value is outside the tabulated data, the formulation uses the maximum and minimum values of the tabulated data accordingly. This approach may create numerical instabilities and errors, particularly in edge cases.

\section{NUMERICAL FORMULATION\label{sec:numerical_formulation}}

The present work uses the {LeMANS} Navier-Stokes solver, a parallel code for unstructured meshes developed at the University of Michigan~\cite{lc_scalabrin_phd_2007}. The code solves the Navier-Stokes equations, including chemical reactions and energy transfer between different energy modes, using a finite volume method with a cell-centered approach. Furthermore, the {LeMANS} solver is capable of handling axisymmetric flow configurations using meshes composed solely of quadrilaterals to better resolve the boundary layers and shock waves~\cite{lc_scalabrin_phd_2007}.

{LEMANS} uses a modified Steger-Warming flux vector splitting (FVS) scheme to discretize the inviscid fluxes across the cell faces~\cite{maccormack_candler_1989}. The modified method switches to the original Steger-Warming FVS scheme, Ref.~\cite{steger_warming_jcp_1981}, in the vicinity of shock waves or other discontinuities by the action of a pressure switch. According to Ref.~\cite{maccormack_candler_1989}, the original Steger-Warming FVS scheme is highly dissipative and suitable for high-gradient regions, such as shock waves. The modified Steger-Warming FVS scheme is less dissipative and suitable for resolving flows within the boundary layer. Moreover, {LeMANS} implements a second-order reconstruction scheme for the inviscid fluxes~\cite{lc_scalabrin_phd_2007}.

The present formulation employs a second-order centered scheme to calculate the viscous fluxes at the cell faces using the property values of the centroid and nodes that compose the face. The values of the properties at the nodes are calculated using a simple average of the cell values of the cells that share the node. This approach increases the stencil employed in the derivative calculations, avoiding loss of accuracy in unstructured meshes. The current formulation employs no-slip and catalytic isothermal wall boundary conditions for the wall-type surfaces for simulations regarding Earth's atmosphere. The present work uses the no-slip and non-catalytic isothermal wall boundary conditions for the Mars Pathfinder test cases. Reference~\cite{fc_moreira_phd_2020} presents a detailed analysis of the influence of catalytic and non-catalytic wall boundary conditions for the configurations used in this work. Reference~\cite{yang_et_al_ijhmt_2020} studies reacting gas-surface interactions for carbon dioxide flows, which provide a good model for applications involving Mars' atmosphere. The axisymmetric source term is spatially discretized with the same approach used for the viscous terms. The present formulation calculates the values of properties on the left and right sides of a cell face using the value of the respective cell centroid and the nodes that compose the control volume~\cite{jawahar_kamath_2000}.

Numerical instabilities may arise related to chemical reactions included in the chemical source term. The first problem that may appear is that the chemical reaction rates may achieve large values depending on the control temperature, \( T_{c} \)\@, especially for the low equilibrium constant values~\cite{c_park_1988}. Another problem that may arise is that the density of the chemical species can assume negative values during the convergence process of the numerical solver. Negative density values yield negative values for the source terms, thus numerical instabilities. Reference~\cite{lc_scalabrin_phd_2007} proposes a modified temperature to overcome the problem related to the source term calculations.

Numerical instabilities may appear due to the use of explicit methods for the time integration of the Navier-Stokes equations with chemical source terms. Moreover, the time step restriction due to the stiffness of the formulation presented does not allow an adequate convergence rate to the solution~\cite{hirsch_2007}. Implicit time integration schemes are better suited for a stiff system of equations, allowing larger time steps while avoiding the growth of numerical instabilities. Therefore, the current formulation uses point and line implicit time integration schemes~\cite{lc_scalabrin_phd_2007,venkatakrishnan_1995}.

\section{RESULTS AND DISCUSSION\label{sec:results_and_discussion}}

This section addresses the results of the hypersonic flow calculation performed for the Earth and Mars atmospheres. The simulations performed for the Earth atmosphere use the {FIRE II} reentry capsule as a test vehicle, while the simulations performed for the Mars atmosphere use the Mars Pathfinder entry capsule as the test vehicle. In both cases, the present work evaluates the impact of the weight factors of Park's two-temperature model control temperature, \( T_{c} \)\@, in the flow behavior in terms of the Mach number distributions, temperature modes distributions, stagnation point convective heat flux, and chemical species mass fractions. The present work uses a set of values, including the often proposed good choices, to assess the impact of the weight factors, which are \( a = 0.4 \)\@, \( 0.5 \)\@, \( 0.6 \)\@, \( 0.7 \)\@, and \( 0.8 \)\@.

\subsection{FIRE II}

\subsubsection{Flow Conditions}

The freestream conditions are based on the experimental data from the {FIRE II} reentry trajectory, described in Table~\ref{tab:fire_2_freestream_conditions}\@. In the referenced table, \( H \) is the altitude relative to the sea level, \( \rho_{\infty} \) is the freestream density, \( T_{\infty} \) is the freestream temperature, \( T_{w} \) is the {FIRE II} wall surface temperature, \( U_{\infty} \) is the flow speed, \( R_{n} \) is the largest curvature radius of the thermal shield, \( R \) is the thermal shield circumference radius used as characteristic length, \( \Mach_{\infty} \) is the freestream Mach number, \( \Reyn_{\infty} \) is the freestream Reynolds number, and \( \Knud_{\infty} \) is the freestream Knudsen number. The freestream gas mixture of Earth's atmosphere is approximately \SI{76.3}{\percent} of \ce{N2} and \SI{23.7}{\percent} of \ce{O2} in terms of mass fraction.

\begin{table}[hbt!]
    \caption{\label{tab:fire_2_freestream_conditions}{FIRE II} freestream conditions.}
    \centering
    \begingroup
    \small
    \begin{tabular}{cccccccccc}
    \hline
        \( H [\unit{\km}] \)
            & \( \rho_{\infty} [\unit[per-mode=symbol]{\kg\per\cubic\m}] \)
            & \( T_{\infty} [\unit{\kelvin}] \)
            & \( T_{w} [\unit{\kelvin}] \)
            & \( U_{\infty} [\unit[per-mode=symbol]{\m\per\s}] \)
            & \( R_{n} [\unit{\m}] \)
            & \( R [\unit{\m}] \)
            & \( \Mach_{\infty} \)
            & \( \Reyn_{\infty} \)
            & \( \Knud_{\infty} \) \\ \hline
        \num{71.02}
            & \num{8.57e-05}
            & \num{210}
            & \num{810}
            & \num{11310}
            & \num{0.935}
            & \num{0.33575}
            & \num{38.85}
            & \num{2.19e+04}
            & \num{2.63e-03} \\
        \num{48.37}
            & \num{1.32e-03}
            & \num{285}
            & \num{1520}
            & \num{9830}
            & \num{0.8052}
            & \num{0.31495}
            & \num{28.99}
            & \num{2.17e+05}
            & \num{1.97e-04} \\
        \num{41.60}
            & \num{3.25e-05}
            & \num{267}
            & \num{503}
            & \num{8100}
            & \num{0.7021}
            & \num{0.29395}
            & \num{24.68}
            & \num{4.32e+05}
            & \num{8.46e-05} \\
    \hline
    \end{tabular}
    \endgroup
\end{table}

The {FIRE II} reentry capsule has three thermal shields. The first thermal shield is the one active for the case of altitude \( H = \SI{71.02}{\km} \) and is ejected at an altitude of approximately \( H = \SI{59.62}{\km} \)\@. The second thermal shield protects the capsule until the altitude of approximately \( H = \SI{48.37}{\km} \)\@. The third and last thermal shield protects the {FIRE II} capsule until landing. Therefore, each case chosen represents one of the different thermal shields of the {FIRE II} capsule.

\subsubsection{Computational Grid}

\begin{figure}[hbt!]
    \centering
    \begin{subfigure}{0.31\linewidth}
       \centering
       \includegraphics[width=\textwidth]{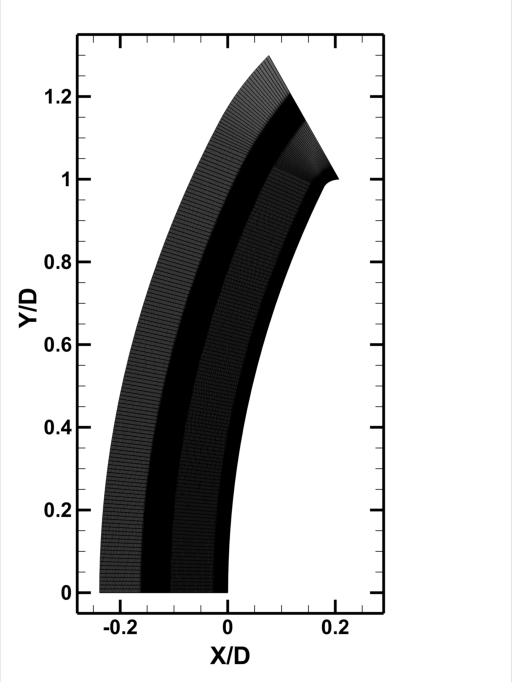}
       \caption{\( H = \SI{71.02}{\km} \)\@.}
       \label{fig:fire_2_7102_mesh}
    \end{subfigure}
    \hspace{0.02\textwidth}
    \begin{subfigure}{0.31\linewidth}
       \centering
       \includegraphics[width=\textwidth]{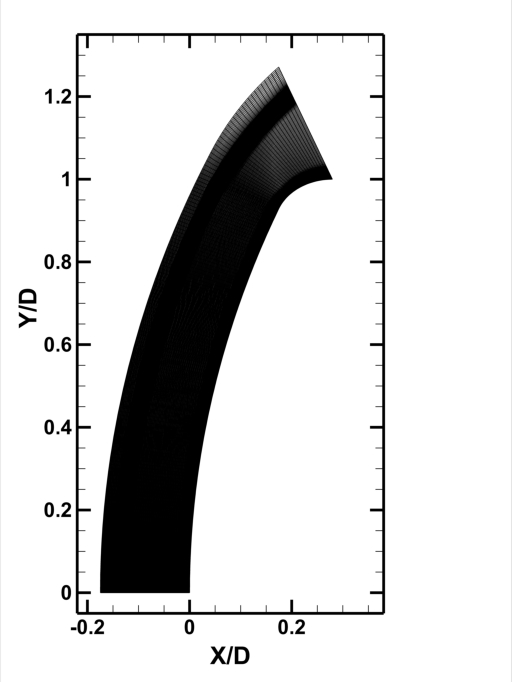}
       \caption{\( H = \SI{48.37}{\km} \)\@.}
       \label{fig:fire_2_4837_mesh}
    \end{subfigure}
    \hspace{0.02\textwidth}
    \begin{subfigure}{0.31\linewidth}
       \centering
       \includegraphics[width=\textwidth]{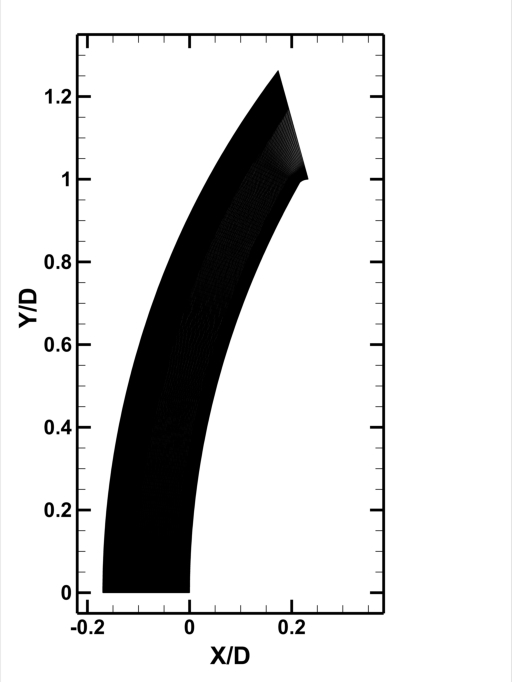}
       \caption{\( H = \SI{41.60}{\km} \)\@.}
       \label{fig:fire_2_4160_mesh}
    \end{subfigure}
    \caption{{FIRE II} computational grids.}
    \label{fig:fire_2_computational_grids}
\end{figure}

\begin{figure}[hbt!]
    \centering
    \begin{subfigure}{0.31\linewidth}
       \centering
       \includegraphics[width=\textwidth]{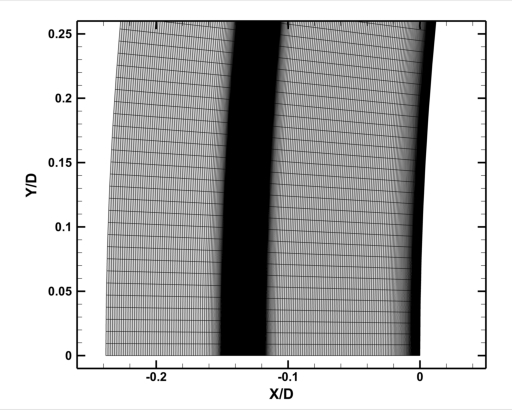}
       \caption{\( H = \SI{71.02}{\km} \)\@.}
       \label{fig:fire_2_7102_mesh_zoomed}
    \end{subfigure}
    \hspace{0.02\textwidth}
    \begin{subfigure}{0.31\linewidth}
       \centering
       \includegraphics[width=\textwidth]{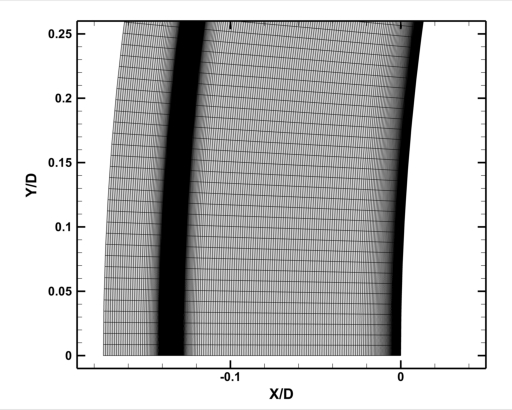}
       \caption{\( H = \SI{48.37}{\km} \)\@.}
       \label{fig:fire_2_4837_mesh_zoomed}
    \end{subfigure}
    \hspace{0.02\textwidth}
    \begin{subfigure}{0.31\linewidth}
       \centering
       \includegraphics[width=\textwidth]{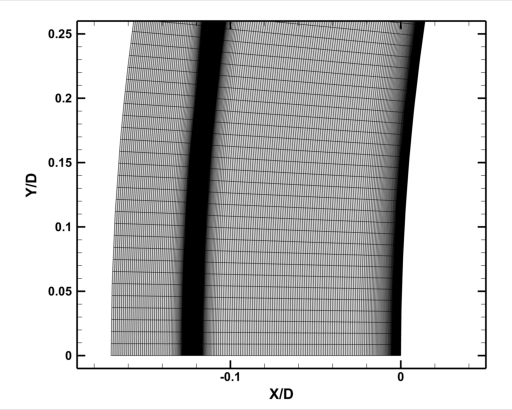}
       \caption{\( H = \SI{41.60}{\km} \)\@.}
       \label{fig:fire_2_4160_mesh_zoomed}
    \end{subfigure}
    \caption{{FIRE II} computational grids zoomed in the stagnation streamline.}
    \label{fig:fire_2_computational_grids_zoomed}
\end{figure}

The grids used for the numerical simulations performed in the present work are composed solely of quadrilaterals in axisymmetric configuration. The computational grids for each thermal shield of the {FIRE II} capsule have two distinct refinement regions. The first refinement region is near the shock wave, where the non-equilibrium phenomena occur with the highest intensity. This region includes the shock wave and the non-equilibrium phenomena immediately behind the shock. The present work has used the experience acquired and the results available in Refs.~\cite{fc_moreira_phd_2020,moreira_wolf_azevedo_scitech_2023} to define this region. Moreover, this work refers to the refinement region, including the shock wave and the non-equilibrium phenomena, as the ``non-equilibrium region.'' The second region of mesh refinement is at the vehicle wall. The mesh refinement at the vehicle wall allows for better capture of wall temperature gradients, thus allowing the correct calculation of the convective heat flux at the vehicle surface. The computational grids used for the  {FIRE II} reentry capsule simulations are presented in Fig.~\ref{fig:fire_2_computational_grids}\@. In those figures, the names of the axes are \( \text{X/D} \) and \( \text{Y/D} \)\@, where \( \text{D} \) is the same as \( R \) in Table~\ref{tab:fire_2_freestream_conditions}\@. Hence, \( \text{X/D} \) and \( \text{Y/D} \) means the same as \( \text{X/R} \) and \( \text{Y/R} \) for the following discussions. Figure~\ref{fig:fire_2_computational_grids_zoomed} presents the computational grids zoomed in the stagnation streamline, allowing better visualization of the mesh refinement approach.

The work refines the mesh regions using a refinement parameter called cell Reynolds number, \( \ReynCell \)\@. The mesh in the non-equilibrium region is uniformly spaced with a stretching factor of \SI{10}{\percent} to create a smooth transition to other regions in the computational grid. Based on the mesh convergence analysis, we identified that \( \ReynCell \approx 5 \) is sufficient to create a mesh with good numerical convergence performance that also yields good results. This study refines the mesh at the vehicle wall by specifying the first and smallest grid distance and applying a stretch factor of \SI{5}{\percent}\@. The smallest grid distance must follow \( \ReynCell < 1 \)\@, as shown in Ref.~\cite{moreira_wolf_azevedo_scitech_2023}. To fulfill this condition, the present work defines the first grid distance at the vehicle wall as \( \Delta n = \SI{5e-07}{\m} \)\@. Table~\ref{tab:fire_2_mesh_config} presents the details of each of the three meshes used in this study. In the table, \( n \) represents the number of cells in the wall-normal and stream-wise direction, respectively.

\begin{table}[hbt!]
    \caption{\label{tab:fire_2_mesh_config}{FIRE II} mesh parameters.}
    \centering
    \begin{tabular}{ccc}
    \hline
        \( H [\unit{\km}] \)
            & \( n_{\text{wall-normal}} \)
            & \( n_{\text{streamwise}} \) \\ \hline
        \num{71.02} & \num{490}  & \num{148} \\
        \num{48.37} & \num{739}  & \num{148} \\
        \num{41.60} & \num{1130} & \num{138} \\
    \hline
    \end{tabular}
\end{table}

\subsubsection{Mach Number and Shock Wave Position}

\begin{figure}[hbt!]
    \centering
    \begin{subfigure}{0.31\linewidth}
       \centering
       \includegraphics[width=\textwidth]{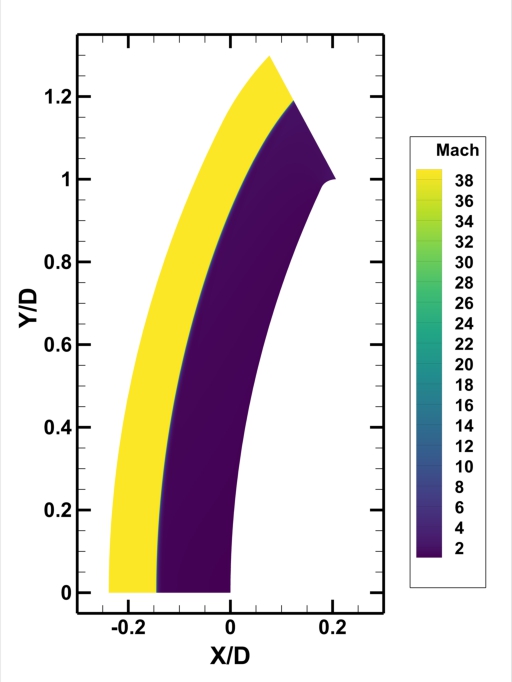}
       \caption{\( H = \SI{71.02}{\km} \)\@.}
       \label{fig:fire_2_7102_mach_contour}
    \end{subfigure}
    \hspace{0.015\textwidth}
    \begin{subfigure}{0.31\linewidth}
       \centering
       \includegraphics[width=\textwidth]{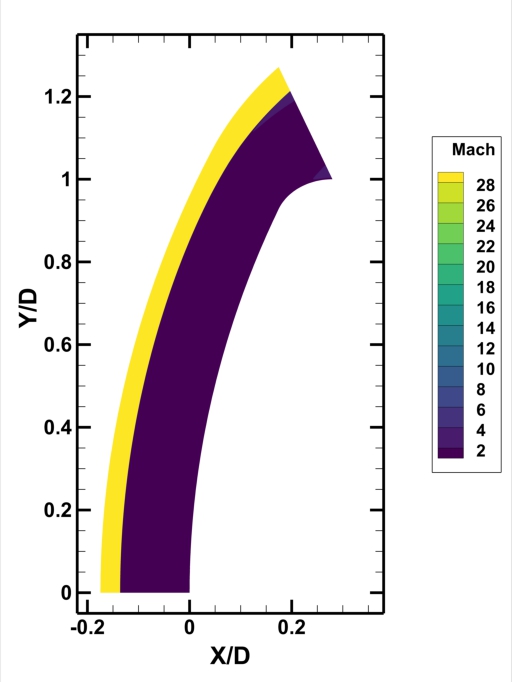}
       \caption{\( H = \SI{48.37}{\km} \)\@.}
       \label{fig:fire_2_4837_mach_contour}
    \end{subfigure}
    \hspace{0.015\textwidth}
    \begin{subfigure}{0.31\linewidth}
       \centering
       \includegraphics[width=\textwidth]{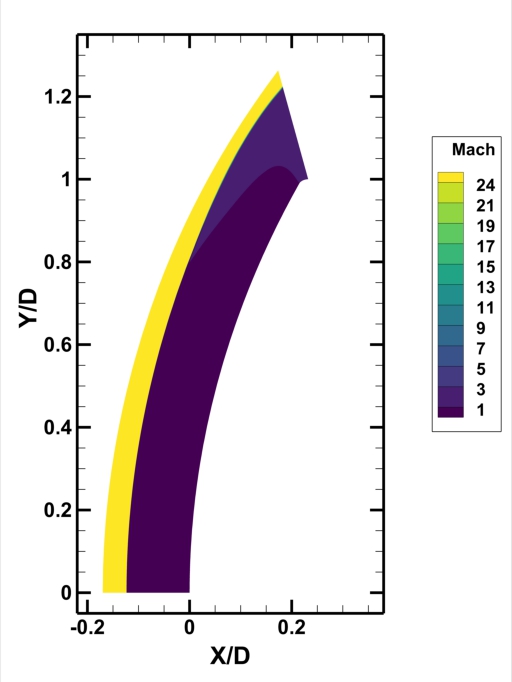}
       \caption{\( H = \SI{41.60}{\km} \)\@.}
       \label{fig:fire_2_4160_mach_contour}
    \end{subfigure}
    \caption{{FIRE II} Mach contours for each computational grid considering \( a = 0.5 \)\@.}
    \label{fig:fire_2_mach_contours}
\end{figure}

Figure~\ref{fig:fire_2_mach_contours} shows the Mach number contours for each thermal shield, considering a weight factor \( a = 0.5 \)\@. The shock standoff distance from the vehicle wall agrees with the results presented by Ref.~\cite{moreira_wolf_azevedo_scitech_2023}\@. The bow shock seems well defined where no intermediate colors are visible. However, this may be due to the scale of the computational domain.

\begin{figure}[hbt!]
    \centering
    \begin{subfigure}{0.41\linewidth}
       \centering
       \includegraphics[width=\textwidth]{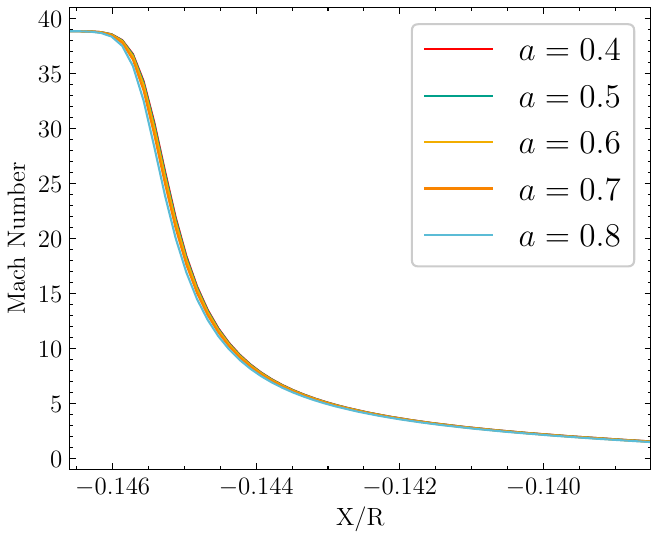}
       \caption{\( H = \SI{71.02}{\km} \)\@.}
       \label{fig:fire_2_7102_mach}
    \end{subfigure}
    \hspace{0.02\textwidth}
    \begin{subfigure}{0.41\linewidth}
       \centering
       \includegraphics[width=\textwidth]{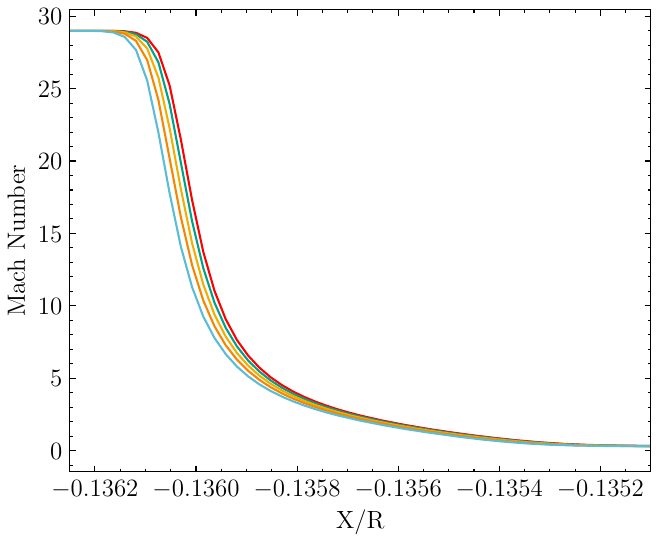}
       \caption{\( H = \SI{48.37}{\km} \)\@.}
       \label{fig:fire_2_4837_mach}
    \end{subfigure}
    \\
    \begin{subfigure}{0.41\linewidth}
       \centering
       \includegraphics[width=\textwidth]{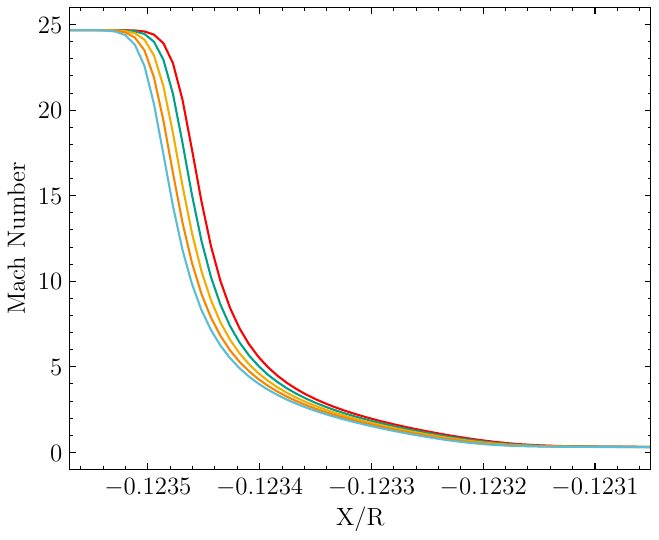}
       \caption{\( H = \SI{41.60}{\km} \)\@.}
       \label{fig:fire_2_4160_mach}
    \end{subfigure}
    \caption{{FIRE II} Mach number distributions along the stagnation streamline in the non-equilibrium region.}
    \label{fig:fire_2_cmp_mach}
\end{figure}

To better visualize the shock structure and standoff distance to the vehicle wall, Fig.~\ref{fig:fire_2_cmp_mach} presents the Mach number distributions along the stagnation streamline in the non-equilibrium region for the weight factor \( a = 0.4 \)\@, \( 0.5 \)\@, \( 0.6 \)\@, \( 0.7 \)\@, and \( 0.8 \)\@. The results presented show that the Mach number distributions decrease rapidly from the freestream Mach number values to almost zero in a small region, as depicted in Fig.~\ref{fig:fire_2_cmp_mach}\@. Moreover, the results show that changes in the weight factor values impact the position of the shock wave. In this case, the results indicate that the shock wave tends to move away from the {FIRE II} wall as the \( a \) weight factor increases. The results in Fig.~\ref{fig:fire_2_cmp_mach} are a consequence of the fact that an increase of the \( a \) weight factor influences the calculation of the control temperature, \( T_{c} \)\@, yielding a temperature distribution closer to the \( T_{tr} \) profile. Since \( T_{tr} \) is typically considerably higher than \( T_{ve} \), the increase of \( a \) changes the shape and location of the maximum value of the control temperature, \( T_{c} \)\@, and its distribution, typically increasing \( T_{c} \) in the non-equilibrium region. These changes affect the calculations of the forward reaction rates, tend to increase the dissociation reactions. Thus, the chemical species mass fraction distributions inside the non-equilibrium region are also going to be affected. However, the changes in the position of the shock waves are small compared to the length of the stagnation streamline.

\subsubsection{Temperature Modes}

\begin{figure}[hbt!]
    \centering
    \begin{subfigure}{0.41\linewidth}
       \centering
       \includegraphics[width=\textwidth]{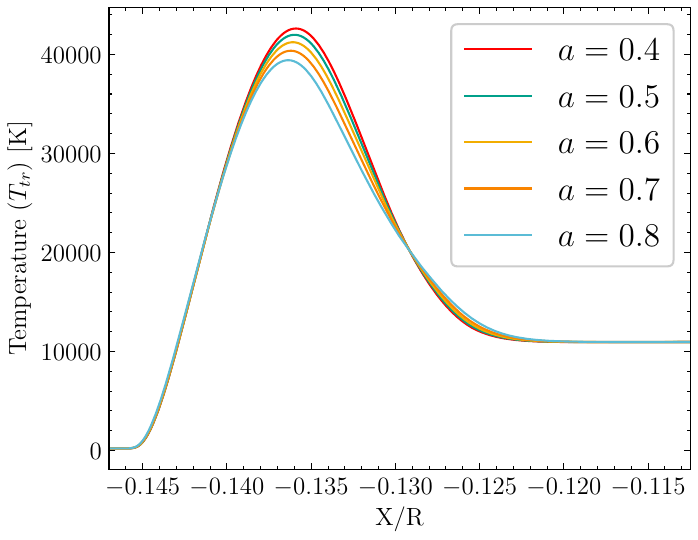}
       \caption{\( H = \SI{71.02}{\km} \)\@.}
       \label{fig:fire_2_7102_t_tr}
    \end{subfigure}
    \hspace{0.02\textwidth}
    \begin{subfigure}{0.41\linewidth}
       \centering
       \includegraphics[width=\textwidth]{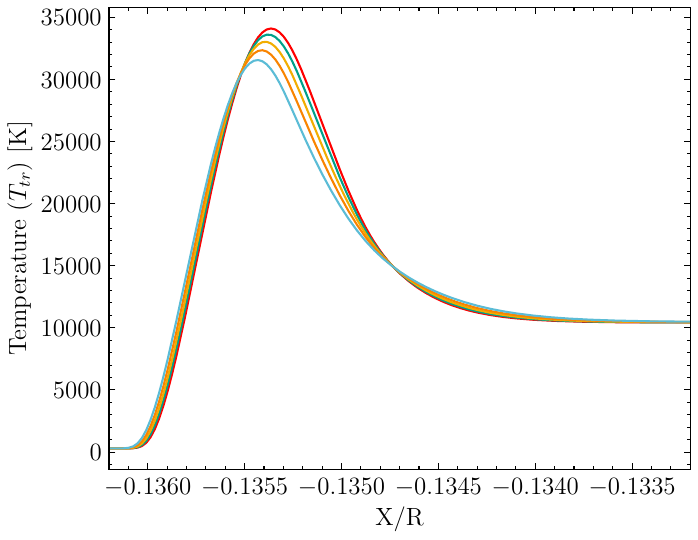}
       \caption{\( H = \SI{48.37}{\km} \)\@.}
       \label{fig:fire_2_4837_t_tr}
    \end{subfigure}
    \\
    \begin{subfigure}{0.41\linewidth}
       \centering
       \includegraphics[width=\textwidth]{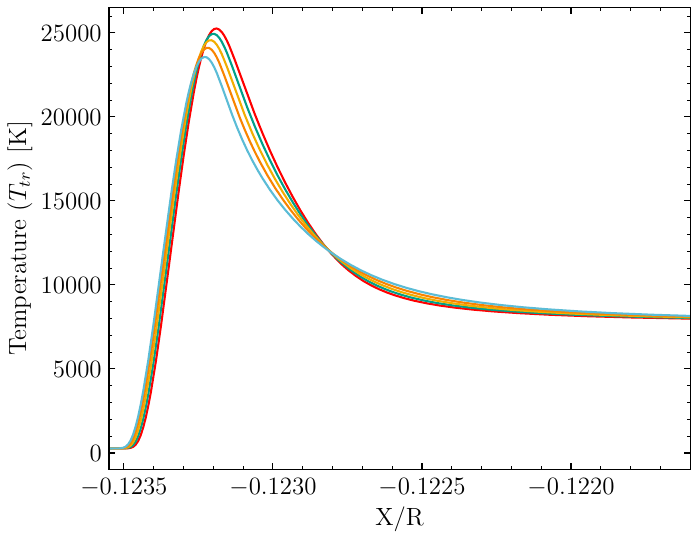}
       \caption{\( H = \SI{41.60}{\km} \)\@.}
       \label{fig:fire_2_4160_t_tr}
    \end{subfigure}
    \caption{{FIRE II} translational-rotational temperature mode distributions along the stagnation streamline in the non-equilibrium region.}
    \label{fig:fire_2_cmp_t_tr}
\end{figure}

Figure~\ref{fig:fire_2_cmp_t_tr} presents the translational-rotational, \( T_{tr} \)\@, temperature mode distributions along the stagnation streamline in the non-equilibrium region for the set of weight factors addressed in this work. The results shown in this figure indicate that the maximum temperature values decrease with an increase in the \( a \) weight factor value. There are also slight changes in the shape and location of the \( T_{tr} \) temperature mode distributions.

The changes in the \( T_{tr} \) distributions for Figs.~\ref{fig:fire_2_4837_t_tr} and \ref{fig:fire_2_4160_t_tr} follow the same behavior observed in the Mach number distributions, Fig.~\ref{fig:fire_2_cmp_mach}\@. In general, the position of the shock wave and the non-equilibrium region agree with numerical data in Refs.~\cite{lc_scalabrin_phd_2007,moreira_wolf_azevedo_scitech_2023,fc_moreira_phd_2020}. However, the present work shows that the \( T_{tr} \) temperature mode distributions in the non-equilibrium region have higher maximum temperature values than those observed in the literature. Reference~\cite{moreira_wolf_azevedo_scitech_2023} predicts maximum values of \( T_{tr} \) closer to the temperature values found in the present work due to a similar approach to refine the mesh near the shock wave and non-equilibrium region. The difference between temperature peaks for \( a = 0.4 \) and \( a = 0.8 \) is around \SI{3200}{\kelvin} in Fig.~\ref{fig:fire_2_7102_t_tr}\@. The same comparison for Figs.~\ref{fig:fire_2_4837_t_tr} and \ref{fig:fire_2_4160_t_tr} yields temperature differences around \SI{2300}{\kelvin} and \SI{1700}{\kelvin}\@, respectively. These results indicate that the temperature variations caused by the weight factors are not small in the non-equilibrium region for the {FIRE II} test cases.

\begin{figure}[hbt!]
    \centering
    \begin{subfigure}{0.41\linewidth}
       \centering
       \includegraphics[width=\textwidth]{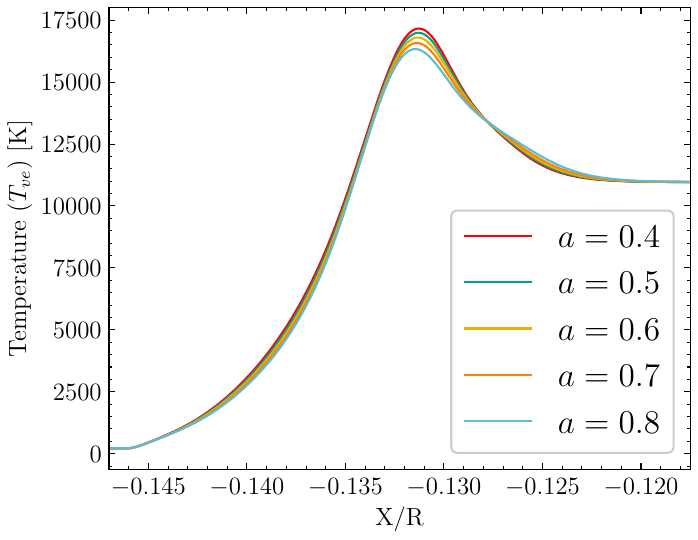}
       \caption{\( H = \SI{71.02}{\km} \)\@.}
       \label{fig:fire_2_7102_t_ve}
    \end{subfigure}
    \hspace{0.02\textwidth}
    \begin{subfigure}{0.41\linewidth}
       \centering
       \includegraphics[width=\textwidth]{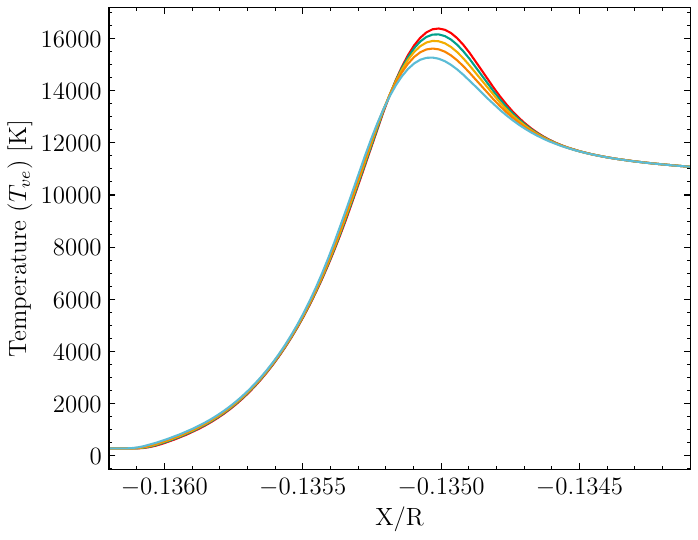}
       \caption{\( H = \SI{48.37}{\km} \)\@.}
       \label{fig:fire_2_4837_t_ve}
    \end{subfigure}
    \\
    \begin{subfigure}{0.41\linewidth}
       \centering
       \includegraphics[width=\textwidth]{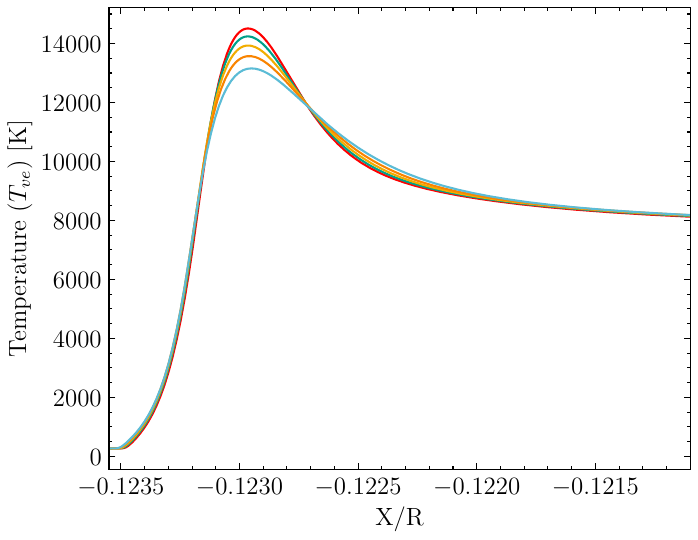}
       \caption{\( H = \SI{41.60}{\km} \)\@.}
       \label{fig:fire_2_4160_t_ve}
    \end{subfigure}
    \caption{{FIRE II} vibrational-electronic temperature mode distributions along the stagnation streamline in the non-equilibrium region.}
    \label{fig:fire_2_cmp_t_ve}
\end{figure}

Figure~\ref{fig:fire_2_cmp_t_ve}presents the vibrational-electronic, \( T_{ve} \)\@, temperature mode distributions along the stagnation streamline in the non-equilibrium region for the set of weight factors addressed in this work. The \( T_{ve} \) temperature mode distributions in Fig.~\ref{fig:fire_2_cmp_t_ve} show similar behavior observed for \( T_{tr} \) distributions in Fig.~\ref{fig:fire_2_cmp_t_tr}, in which the increase the \( a \) weight factor yields lower values of temperature peaks in the non-equilibrium region. Different from the \( T_{tr} \) temperature mode distributions, the \( T_{ve} \) temperature mode distributions show no clear difference in the position of the curves when varying the weight factors. Moreover, the predicted maximum temperature peaks for the \( T_{ve} \) temperature mode are also higher than the values predicted in the literature~\cite{lc_scalabrin_phd_2007,moreira_wolf_azevedo_scitech_2023,fc_moreira_phd_2020}. The difference between temperature peaks for \( a = 0.4 \) and \( a = 0.8 \) is around \SI{800}{\kelvin} in Fig.~\ref{fig:fire_2_7102_t_ve}\@.

The translational-rotational, \( T_{tr}\)\@, temperature mode have temperature values higher than the vi\-bra\-tion\-al-electronic, \( T_{ve} \)\@, temperature mode in the non-equilibrium region. This behavior relates to the fact that the \( T_{tr} \) temperature mode requires lower energy values to excite one energy level when compared to the \( T_{ve} \) temperature mode. Therefore, increasing the \( a \) weight factor yields a control temperature, \( T_{c} \)\@, mainly governed by the \( T_{tr} \) temperature mode.

Reference~\cite{niu_et_al_2018} compares the impact of different temperature models, including Park's two-temperature model, for reactive hypersonic flows in non-equilibrium conditions. Reference~\cite{niu_et_al_2018} observes the same behavior identified in the present work for the vibrational-electronic temperature mode when comparing \( a = 0.5 \) and \( a = 0.7 \) weight factors. However, Ref.~\cite{niu_et_al_2018} does not observe differences regarding the \( T_{tr} \) temperature mode distributions when changing the weight factor values. This result is in contrast with the behavior observed in Fig.~\ref{fig:fire_2_cmp_t_tr}\@, which shows that varying the weight factor values impacts the \( T_{tr} \) distributions.

The differences in the results of the present work and those in Ref.~\cite{niu_et_al_2018} could be related to geometry and freestream conditions. Reference~\cite{niu_et_al_2018} simulated hypersonic flows over the {BUSV} and {RAM-C II} blunt body geometries. Moreover, the freestream conditions are different. Reference~\cite{fc_moreira_phd_2020} used the same numerical tool, {LeMANS}, and obtained results similar to Ref.~\cite{niu_et_al_2018} for \( a = 0.5 \) and \( a = 0.7 \)\@. Therefore, it is plausible to state that the conditions used for the {FIRE II} simulation performed in this work may be sufficiently harsh to show the impact of the weight factors in the \( T_{tr} \) temperature mode distributions.

The change in the weight factor values does not impact the flow regions in the equilibrium state. These regions are the freestream flow and the flow between the non-equilibrium and near-wall regions. According to the literature, the different temperature modes assume the same value in the equilibrium regions. The flow region near the vehicle wall presents a non-equilibrium behavior. However, the non-equilibrium intensity is small and does not significantly cause a visible impact on distributions of properties, as shown in Ref.~\cite{gm_poltronieri_msc_2024}. Therefore, the present work does not include any specific analysis of the equilibrium and near-wall regions.

\subsubsection{Stagnation Point Convective Heat Flux}

\begin{figure}[hbt!]
    \centering
    \begin{subfigure}{0.31\linewidth}
       \centering
       \includegraphics[width=\textwidth]{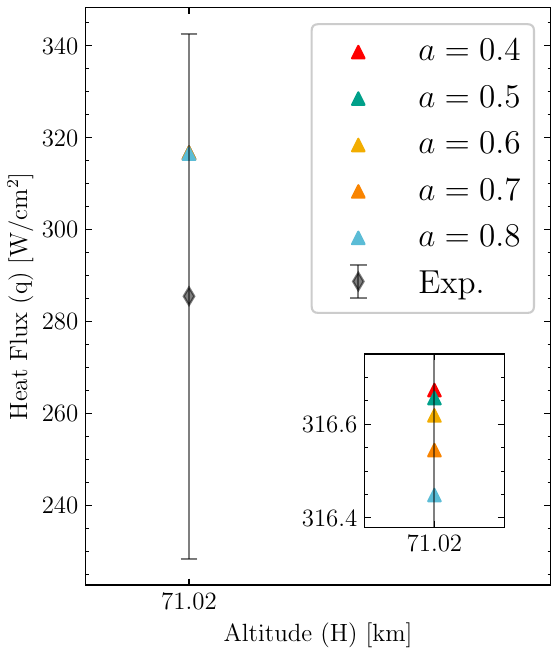}
       \caption{\( H = \SI{71.02}{\km} \)\@.}
       \label{fig:fire_2_7102_q}
    \end{subfigure}
    \hspace{0.02\textwidth}
    \begin{subfigure}{0.31\linewidth}
       \centering
       \includegraphics[width=\textwidth]{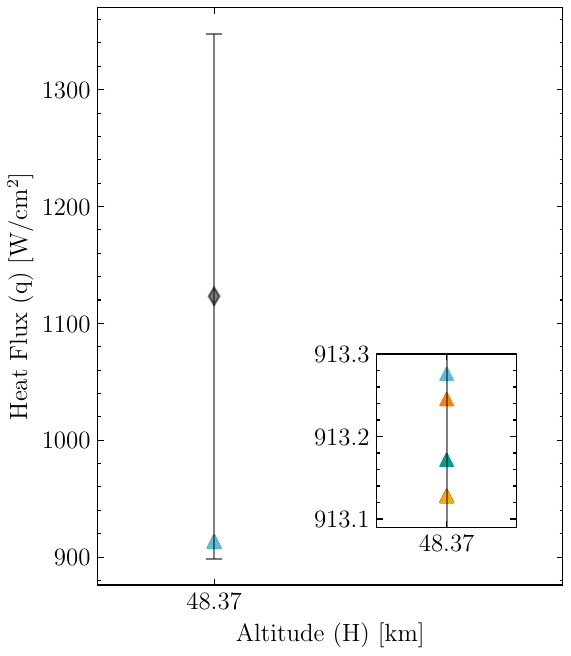}
       \caption{\( H = \SI{48.37}{\km} \)\@.}
       \label{fig:fire_2_4837_q}
    \end{subfigure}
    \hspace{0.02\textwidth}
    \begin{subfigure}{0.31\linewidth}
       \centering
       \includegraphics[width=\textwidth]{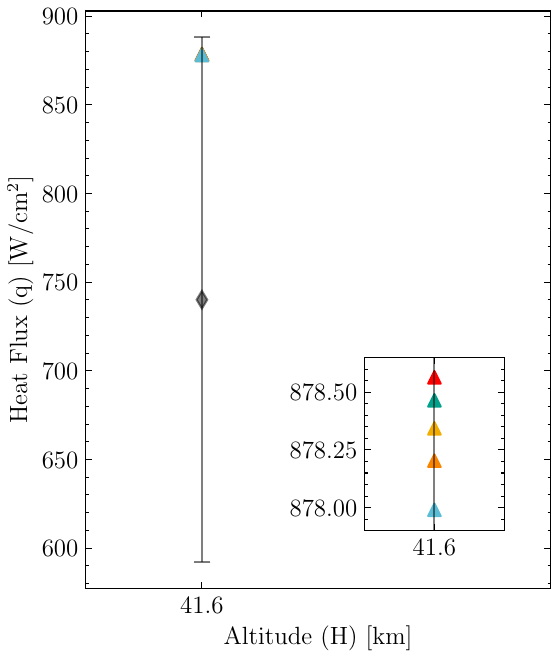}
       \caption{\( H = \SI{41.60}{\km} \)\@.}
       \label{fig:fire_2_4160_q}
    \end{subfigure}
    \caption{{FIRE II} stagnation point convective heat flux.}
    \label{fig:fire_2_cmp_q}
\end{figure}

Figure~\ref{fig:fire_2_cmp_q} shows the stagnation convective heat flux for each case presented for the {FIRE II} and each set of weight factors proposed in this work. This figure also includes experimental data for the stagnation point heat flux from Ref.~\cite{cornette_1966}. The experimental data used as a reference represents the total heat flux at the stagnation point, considering both convective and radiative portions. Moreover, the experimental data has a margin of error of \( \pm \SI{20}{\percent} \)\@, represented by the black diamond and error bars in Fig.~\ref{fig:fire_2_cmp_q}\@.

The stagnation point convective heat fluxes predicted in this work are within the error margin of the experimental data. The variation of the weight factor values causes small differences in the stagnation point convective heat flux values obtained in the simulations. The differences between the highest and lowest values of stagnation point convective heat flux predicted are, at maximum, \SI{0.9}{\percent} of the average values of the stagnation point convective heat flux simulated for each case. However, the numerical tool used in the present work is not capable of coupling radiative phenomena to assess the impact of the weight factor on the total stagnation point heat flux. Data on the literature indicate that the radiative portion of the total stagnation point heat flux may be affected by the chemical species composing the flow~\cite{boyd_schwartzentruber_2017}.

\subsubsection{Chemical Species Mass Fraction}

\begin{figure}[hbt!]
    \centering
    \begin{subfigure}{0.41\linewidth}
       \centering
       \includegraphics[width=\textwidth]{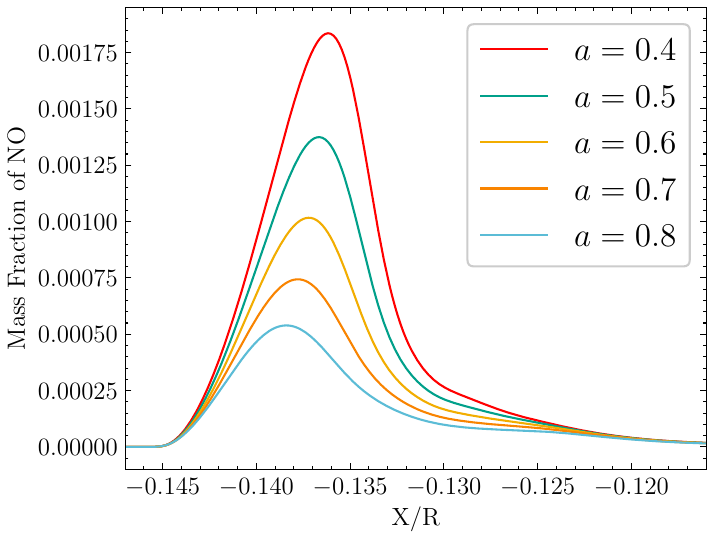}
       \caption{\( H = \SI{71.02}{\km} \)\@.}
       \label{fig:fire_2_7102_chemical_species_NO}
    \end{subfigure}
    \hspace{0.02\textwidth}
    \begin{subfigure}{0.41\linewidth}
       \centering
       \includegraphics[width=\textwidth]{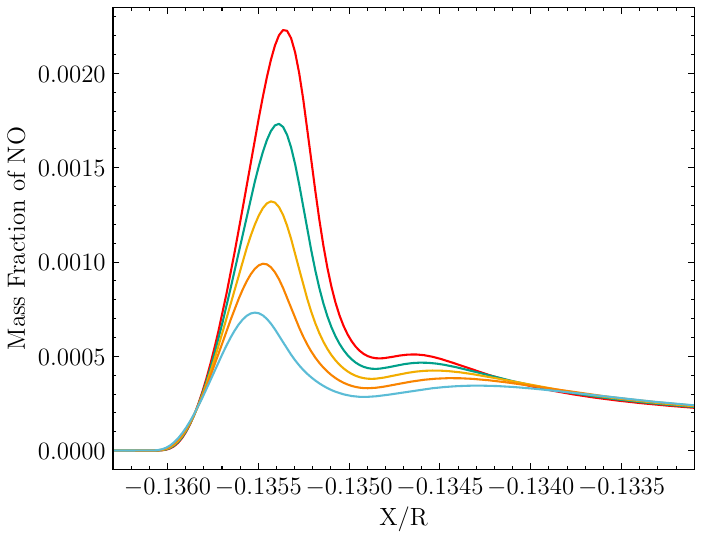}
       \caption{\( H = \SI{48.37}{\km} \)\@.}
       \label{fig:fire_2_4837_chemical_species_NO}
    \end{subfigure}
    \\
    \begin{subfigure}{0.41\linewidth}
       \centering
       \includegraphics[width=\textwidth]{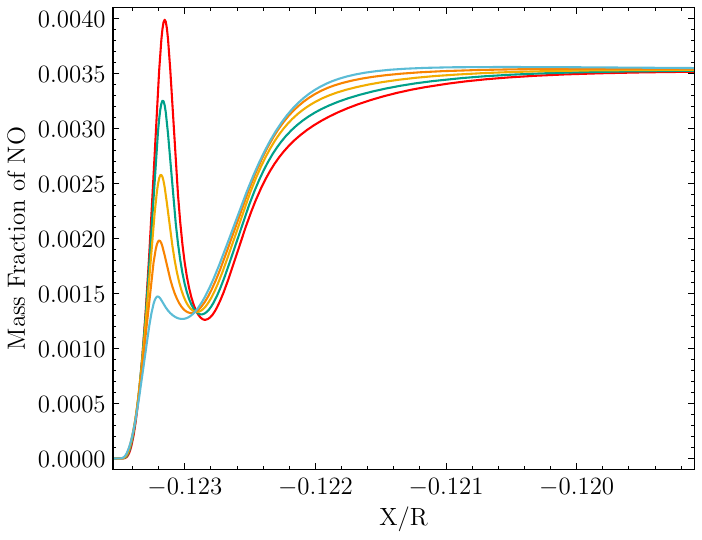}
       \caption{\( H = \SI{41.60}{\km} \)\@.}
       \label{fig:fire_2_4160_chemical_species_NO}
    \end{subfigure}
    \caption{\ce{NO} mass fraction distributions along the stagnation streamline in the non-equilibrium region for the {FIRE II} capsule.}
    \label{fig:fire_2_cmp_chemical_species_NO}
\end{figure}

As mentioned, the present study models Earth's atmosphere by an 11-species chemical model and considers three different test cases for the {FIRE II} reentry vehicle. Thus, presenting all the data gathered for the distributions of mass fractions for each chemical species is not feasible. Therefore, the present study will focus on presenting the mass fraction distributions of the chemical species that showed the largest changes by varying Park's two-temperature model weight factors. Moreover, the chosen chemical species are also strong radiators~\cite{boyd_schwartzentruber_2017,priyadarshini_et_al_aiaa_jthtc_2018}. Therefore, this work presents data for the chemical species \ce{N} and \ce{NO}\@. Figure~\ref{fig:fire_2_cmp_chemical_species_NO} shows the \ce{N} mass fraction distributions along the stagnation streamline in the non-equilibrium region for the set of weight factors \( a = 0.4 \)\@, \( 0.5 \)\@, \( 0.7 \)\@, \( 0.8 \)\@, and \( 0.8 \)\@.

\begin{figure}[hbt!]
    \centering
    \begin{subfigure}{0.41\linewidth}
       \centering
       \includegraphics[width=\textwidth]{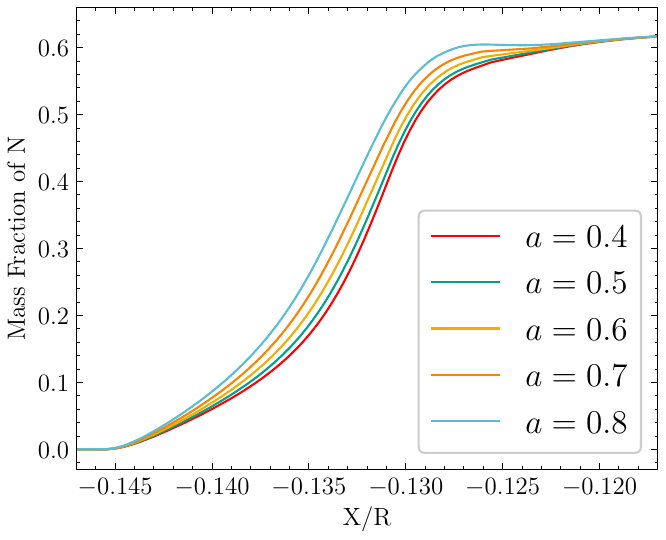}
       \caption{\( H = \SI{71.02}{\km} \)\@.}
       \label{fig:fire_2_7102_chemical_species_N}
    \end{subfigure}
    \hspace{0.02\textwidth}
    \begin{subfigure}{0.41\linewidth}
       \centering
       \includegraphics[width=\textwidth]{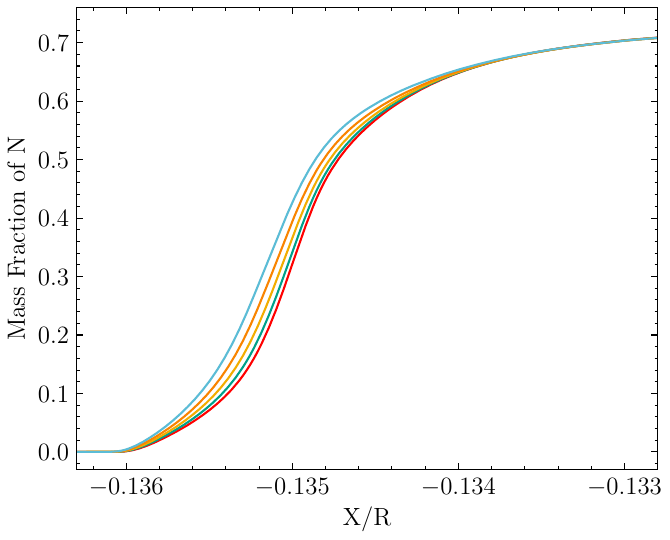}
       \caption{\( H = \SI{48.37}{\km} \)\@.}
       \label{fig:fire_2_4837_chemical_species_N}
    \end{subfigure}
    \\
    \begin{subfigure}{0.41\linewidth}
       \centering
       \includegraphics[width=\textwidth]{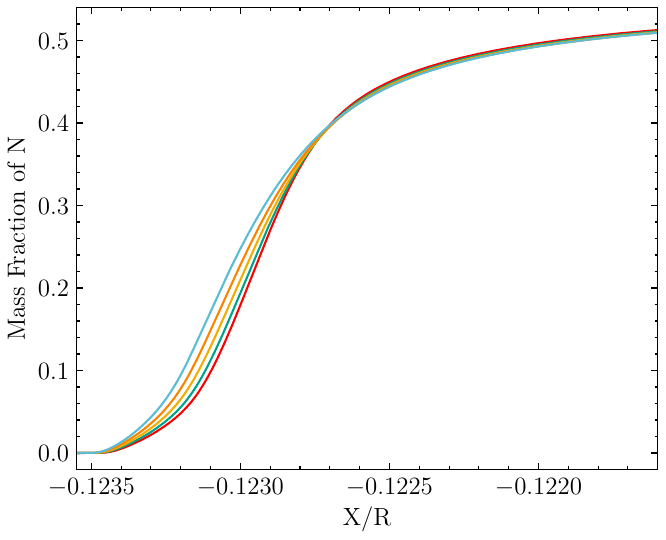}
       \caption{\( H = \SI{41.60}{\km} \)\@.}
       \label{fig:fire_2_4160_chemical_species_N}
    \end{subfigure}
    \caption{\ce{N} mass fraction distributions along the stagnation streamline in the non-equilibrium region for the {FIRE II} capsule.}
    \label{fig:fire_2_cmp_chemical_species_N}
\end{figure}

Figure~\ref{fig:fire_2_cmp_chemical_species_N} shows that the increase of the \( a \) weight factor accelerates the production of \ce{N}\@, which agrees with the expected chemical behavior. The \ce{N} is the product of the \ce{N2} dissociation reaction,
\begin{equation}
    \ce{N2 + M <=> 2N + M}
    \qquad \text{,}
\end{equation}
which is an endothermic reaction. The increase in the overall value of the control temperature, \( T_{c} \)\@, distributions, caused by the increased value of the \( a \) weight factor, favors the endothermic reactions. Therefore, it favors the production of \ce{N}\@. The lines of the distributions seem evenly spaced, indicating that there is no ``convergence'' to a single set of weight factor values. Analyzing Fig.~\ref{fig:fire_2_4160_chemical_species_N} closely in the region \( \text{X/R} > -0.1227 \)\@, it is possible to observe an inversion in the influence of the weight factors' value. The lower the \( a \) weight factor values, the higher the magnitude of the mass fraction distributions of \ce{N} for this particular region.

The behavior observed for the region \( \text{X/R} > -0.1227 \) is unintuitive, and it also relates to the control temperature, \( T_{c} \)\@. In this particular condition, the \( T_{ve} \) temperature mode distributions assume values higher than the \( T_{tr} \) temperature mode distributions because of its ``inertia'' to relax to the equilibrium state. This condition allows for the control temperature, \( T_{c} \)\@, to assume higher temperature distributions for lower values of the \( a \) weight factor for a small part of the non-equilibrium region. Note that the differences caused by the change in the weight factor values have a consistent gap between them, and it seems that they do not ``converge'' to a unique solution within the non-equilibrium region. Continuing the analysis on the chemical species mass fraction distributions, Fig.~\ref{fig:fire_2_cmp_chemical_species_NO} presents the \ce{NO} mass fraction distributions along the stagnation streamline in the non-equilibrium region for the set of weight factors \( a = 0.4 \)\@, \( 0.5 \)\@, \( 0.7 \)\@, \( 0.8 \)\@, and \( 0.8 \)\@.

The results in Fig.~\ref{fig:fire_2_cmp_chemical_species_NO} indicate that the \ce{NO} is produced and consumed almost entirely in the non-equilibrium region of the case of altitude \( H = \SI{71.02}{\km} \)\@, Fig.~\ref{fig:fire_2_7102_chemical_species_NO}\@, where the same does not occur in lower altitudes, Figs.~\ref{fig:fire_2_4837_chemical_species_NO} and \ref{fig:fire_2_4160_chemical_species_NO}\@. This behavior relates to the shock layer temperature distributions. Depending on the shock layer temperature, chemical reactions with \ce{NO} as reactant may be favored, such as the dissociation of \ce{NO}\@. The temperature mode distribution analysis indicates that cases with higher altitudes and flow speeds yield higher temperature mode profiles in the shock layer. Figure~\ref{fig:fire_2_4160_chemical_species_NO} shows a different behavior than the other two cases considered for the {FIRE II} reentry capsule. The \ce{NO} mass fraction distributions have a first peak closer to the freestream flow. Then, there is a local minimum of \ce{NO} mass fraction value, different for each set of weight factors. For \( \text{X/R} \approx -0.1227 \) and larger, the \ce{NO} mass fraction distributions increase in value until the shock layer equilibrium condition, defined by the shock layer temperature.

The increase of the \( a \) weight factor accelerates the consumption of \ce{NO} by favoring the dissociation reaction of \ce{NO}\@. This behavior governs the peaks of \ce{NO} mass fraction observed in the mass fraction distributions presented in Fig.~\ref{fig:fire_2_cmp_chemical_species_NO}\@. The increase of the \( a \) weight factor yields, in general, lower magnitude \ce{NO} mass fraction distributions. As mentioned, Fig.~\ref{fig:fire_2_4160_chemical_species_NO} shows a different behavior for a small length within the non-equilibrium region near the transition to the shock layer region. In this region, lower values of the \( a \) weight factor ``delay'' the production of \ce{NO} towards the shock layer. This ``inversion'' in the expected behavior occurs in a similar area as discussed previously for the \ce{N} chemical species. The control temperature, \( T_{c} \)\@, calculated in this area, has temperature distributions with higher magnitude for lower values of the \( a \) weight factor. The changes caused by varying the weight factors are significant for the \ce{NO} mass fraction distributions within the non-equilibrium region. Moreover, the different sets of weight factors do not ``converge'' to a unique \ce{NO} mass fraction distribution. Instead, the changes in the weight factors cause the solution to produce significantly different mass fraction distributions. The other chemical species considered in the chemical model for Earth's atmosphere also showed behavior following the same physicochemical principles discussed here. The chemical species mass fraction distributions vary with the weight factor changes, a consequence of the control temperature calculation and the type of chemical reactions.

\subsection{Mars Pathfinder}

\subsubsection{Flow Conditions}

This work considers two types of freestream flow gas mixtures to simulate Mars' atmosphere. One of the mixtures is composed, at freestream conditions, solely of carbon dioxide, \ce{CO2}\@, while the other is a mixture of \SI{95}{\percent} of \ce{CO2} and \SI{5}{\percent} of molecular nitrogen, \ce{N2}\@. The percentages are based on the mass fraction. The freestream conditions mimic the experimental data available for the Mars Pathfinder capsule, obtained in the {HYPULSE} expansion tube for carbon dioxide flows. Table~\ref{tab:mars_pathfinder_freestream_conditions} presents the freestream conditions used for the Mars Pathfinder hypersonic flow simulations. The term ``Run 749'' refers to a particular experiment reported in Ref.~\cite{hollis_1996}\@.

\begin{table}[!htpb]
    \caption{\label{tab:mars_pathfinder_freestream_conditions} Mars Pathfinder ``Run 749'' freestream conditions.}
    \centering
    \begin{tabular}{ccccccccc}
    \hline
        Gas Mixture
            & \( \rho_{\infty} [\unit[per-mode=symbol]{\kg\per\cubic\m}] \)
            & \( T_{\infty} [\unit{\kelvin}] \)
            & \( T_{w} [\unit{\kelvin}] \)
            & \( U_{\infty} [\unit[per-mode=symbol]{\m\per\s}] \)
            & \( R [\unit{\m}] \)
            & \( \Mach_{\infty} \)
            & \( \Reyn_{\infty} \)
            & \( \Knud_{\infty} \) \\ \hline
        \ce{CO2}
            & \num{5.75e-03}
            & \num{1045}
            & \num{300}
            & \num{4769}
            & \num{0.0254}
            & \num{9.89}
            & \num{2.03e+04}
            & \num{6.63e-04} \\
        \ce{CO2 + N2}
            & \num{5.75e-03}
            & \num{1045}
            & \num{300}
            & \num{4769}
            & \num{0.0254}
            & \num{9.89}
            & \num{2.03e+04}
            & \num{6.64e-04} \\
    \hline
    \end{tabular}
\end{table}

In the table, \( \rho_{\infty} \) is the freestream density, \( T_{\infty} \) is the freestream temperature, \( T_{w} \) is the Mars Pathfinder capsule wall surface temperature, \( U_{\infty} \) is the flow speed, \( R \) is the thermal shield circumference radius used as characteristic length, \( \Mach_{\infty} \) is the freestream Mach number, \( \Reyn_{\infty} \) is the freestream Reynolds number, and \( \Knud_{\infty} \) is the freestream Knudsen number. Moreover, the {HYPULSE} experiment was only conducted for \ce{CO2} flows. However, Mars' atmosphere composition is, approximately, \SI{95}{\percent} of \ce{CO2}\@, \SI{3}{\percent} of \ce{N2}\@, and \SI{2}{\percent} of \ce{Ar}\@. Therefore, the assumption of \SI{95}{\percent} of \ce{CO2} and \SI{5}{\percent} of \ce{N2} for the Mars atmosphere gas mixture composition is not unreasonable.

\subsubsection{Computational Grids}

The computational grids used for the numerical simulations performed in the present work are also composed solely of quadrilaterals in axisymmetric configuration. Similar to the computational grids for the {FIRE II} capsule, the meshes used for the Mars Pathfinder test case have two different refinement regions. The first refinement region includes the shock wave and the non-equilibrium phenomena and is defined by two internal grid lines. The second refinement region is at the vehicle wall surface. Likewise, the first refinement region is called the ``non-equilibrium region'' and aims to better resolve the shock wave and property gradients related to non-equilibrium phenomena. The refinement region near the vehicle wall allows better capture of temperature gradients, aiming for the correct calculation of the wall convective heat flux.

\begin{figure}[hbt!]
    \centering
    \begin{subfigure}{0.3\linewidth}
       \centering
       \includegraphics[width=\textwidth]{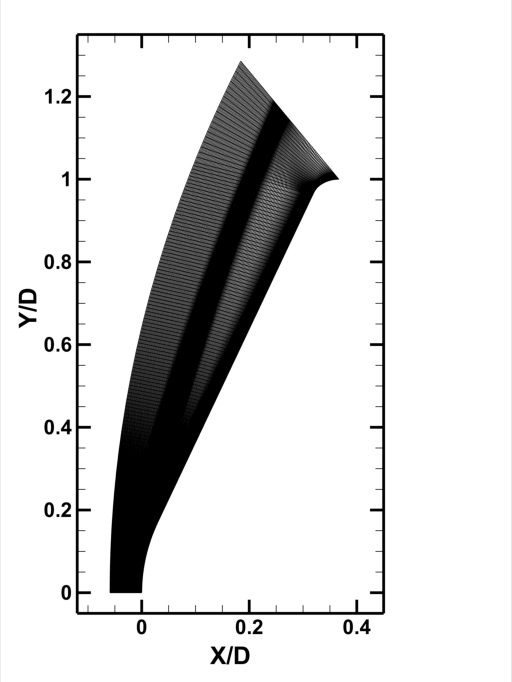}
       \caption{Computational grid.}
       \label{fig:mars_full_grid}
    \end{subfigure}
    \hspace{0.02\textwidth}
    \begin{subfigure}{0.45\linewidth}
       \centering
       \includegraphics[width=\textwidth]{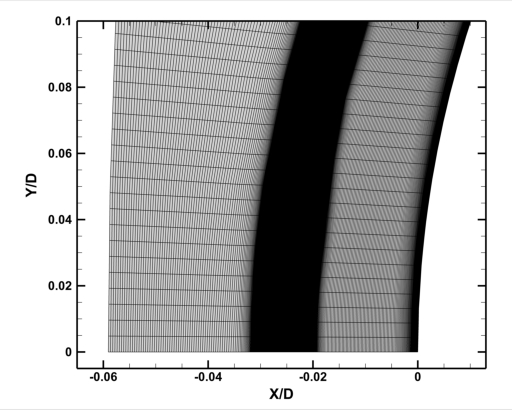}
       \caption{Mesh zoomed in the stagnation streamline.}
       \label{fig:mars_zoomed_grid}
    \end{subfigure}
    \caption{Mars Pathfinder computational grid.}
    \label{fig:mars_computational_grid}
\end{figure}

The definition of the non-equilibrium region follows previous results in Refs.~\cite{moreira_wolf_azevedo_scitech_2021,moreira_wolf_azevedo_jhmt_2021}\@. Moreover, the non-equilibrium refinement region is uniformly spaced and based on the parameter \( \ReynCell \approx 1 \)\@, where a stretching factor of \SI{10}{\percent} guarantees a smooth transition to other mesh regions. This study refines the mesh at the vehicle wall with the same properties as the grid for the {FIRE II} capsule, where the smallest grid distance is \( \Delta n = \SI{5e-07}{\m} \) and the stretching factor is \SI{5}{\percent}\@. The mesh used in the present work simulation regarding the Mars Pathfinder capsule has 160 cells in the streamwise direction and 460 cells in the wall-normal direction. Figure~\ref{fig:mars_computational_grid} shows the Mars Pathfinder computational grid used for the calculation of the results presented here.

A mesh convergence analysis performed in the context of the present effort indicated that the \( \ReynCell \approx 5 \) is not sufficient to allow the correct refinement of the non-equilibrium region. This result differs from the analysis performed for the {FIRE II} computational grids. The mesh that yielded good property distributions was the one with \( \ReynCell \approx 1 \) in the non-equilibrium region. Reference~\cite{gm_poltronieri_msc_2024} provides more details about the mesh convergence analysis performed for the Mars Pathfinder capsule.

\subsubsection{Mach Number and Shock Wave Position}

\begin{figure}[hbt!]
    \centering
    \begin{subfigure}{0.41\linewidth}
       \centering
       \includegraphics[width=0.7\textwidth]{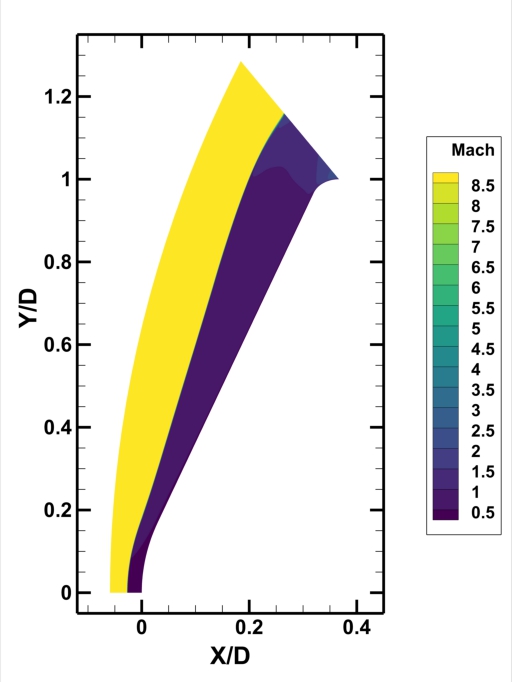}
       \caption{\ce{CO2} flow.}
       \label{fig:mars_cmp_mach_contour_co2}
    \end{subfigure}
    \hspace{0.02\textwidth}
    \begin{subfigure}{0.41\linewidth}
       \centering
       \includegraphics[width=0.7\textwidth]{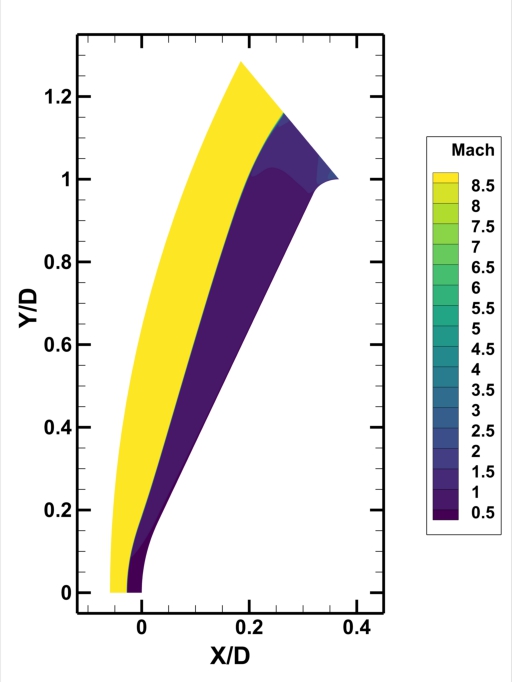}
       \caption{\ce{CO2 + N2} flow.}
       \label{fig:mars_cmp_mach_contour_co2_n2}
    \end{subfigure}
    \caption{Mars Pathfinder Mach number distributions along the stagnation streamline in the non-equilibrium region.}
    \label{fig:mars_cmp_mach_contour}
\end{figure}

Figure~\ref{fig:mars_cmp_mach_contour} shows the Mach number contours for each Mars Pathfinder case considering \( a = 0.5 \)\@. The shock standoff distance from the vehicle wall agrees with the results in Refs.~\cite{moreira_wolf_azevedo_scitech_2021,moreira_wolf_azevedo_jhmt_2021} The bow shock seems well defined where no intermediate color are visible. However, the absence of intermediate colors may be due to the scale of the computational domain compared to the scale of the Mach number distribution changes. Hence, zooming in on the non-equilibrium region along the stagnation streamline allows better visualization of the shock structure. In addition to the Mach number results already presented, this work also included the results for the Mach number distributions along the stagnation streamline for all sets of weight factor chosen by the present work.

\begin{figure}[hbt!]
    \centering
    \begin{subfigure}{0.41\linewidth}
       \centering
       \includegraphics[width=0.8\textwidth]{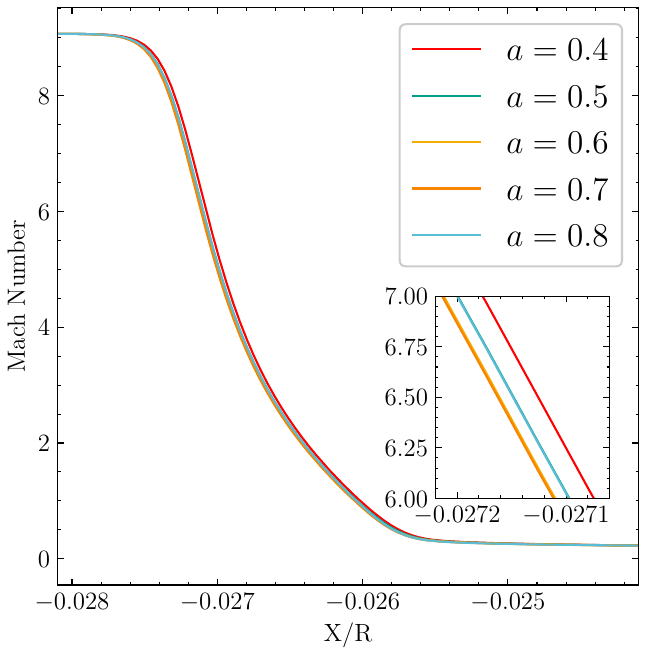}
       \caption{\ce{CO2} flow.}
       \label{fig:mars_cmp_mach_co2}
    \end{subfigure}
    \hspace{0.02\textwidth}
    \begin{subfigure}{0.41\linewidth}
       \centering
       \includegraphics[width=0.8\textwidth]{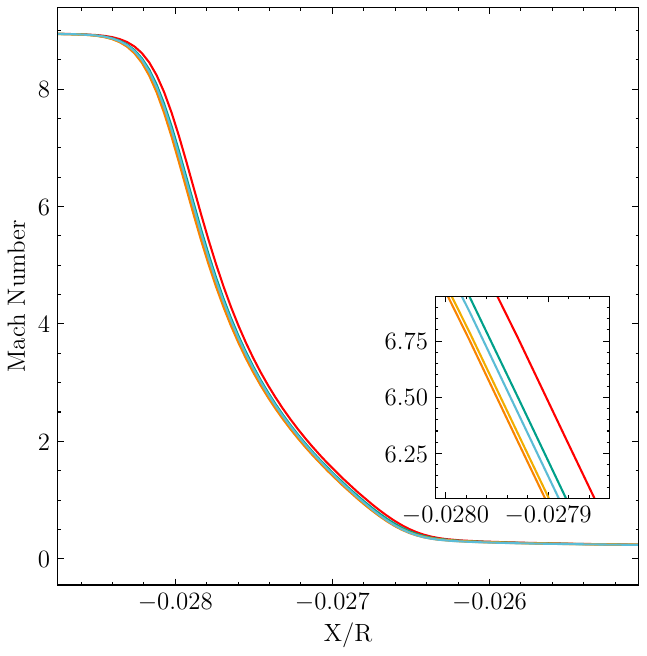}
       \caption{\ce{CO2 + N2} flow.}
       \label{fig:mars_cmp_mach_co2_n2}
    \end{subfigure}
    \caption{Mars Pathfinder Mach number distributions along the stagnation streamline in the non-equilibrium region.}
    \label{fig:mars_cmp_mach}
\end{figure}

Therefore, Figure~\ref{fig:mars_cmp_mach} shows the Mach number distributions along the stagnation streamline in the non-equilibrium region for each set of weight factors proposed in the present work. Figure~\ref{fig:mars_cmp_mach_co2} represents the Mach number distributions for the flow composed solely of \ce{CO2} while Fig.~\ref{fig:mars_cmp_mach_co2_n2} represents results for the gas mixture of \ce{CO2 + N2}\@. The results indicate a rapid decrease in the Mach number values from the freestream condition to almost zero. The results obtained for the shock wave position in this analysis are in good agreement with the numerical data in Refs.~\cite{moreira_wolf_azevedo_scitech_2021,moreira_wolf_azevedo_jhmt_2021}. The position of the shock wave seems to move away from the vehicle body as the \( a \) weight factor value increases. This behavior is similar to the one observed in the {FIRE II} results regarding the Mach number distributions. However, this trend breaks down for \( a = 0.7 \) and \( a = 0.8 \)\@, where the latter predicts a shock wave closer to the vehicle body.

The magnitude of the movement of the shock wave standoff position is significantly small when compared to the length of the stagnation streamline. The change in the behavior of the shock wave position with the increase of the \( a \) weight factor can arise from numerical fluctuations during the solution process. However, more analyses are needed to fully understand the root cause of the modification of the trend in the shock motion with the increase of the \( a \) factor. Furthermore, it is also possible that the flow conditions allow for such reversion in the trend of the shock wave position.

\subsubsection{Temperature Modes}

\begin{figure}[hbt!]
    \centering
    \begin{subfigure}{0.41\linewidth}
       \centering
       \includegraphics[width=\textwidth]{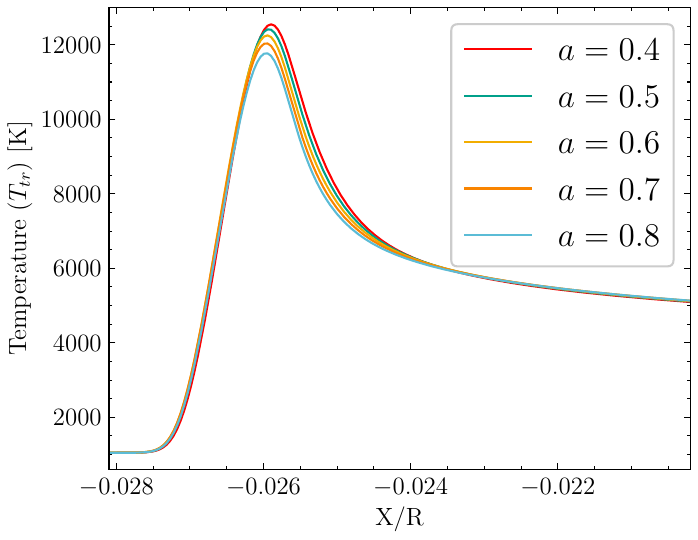}
       \caption{\ce{CO2} flow.}
       \label{fig:mars_cmp_t_tr_co2}
    \end{subfigure}
    \hspace{0.02\textwidth}
    \begin{subfigure}{0.41\linewidth}
       \centering
       \includegraphics[width=\textwidth]{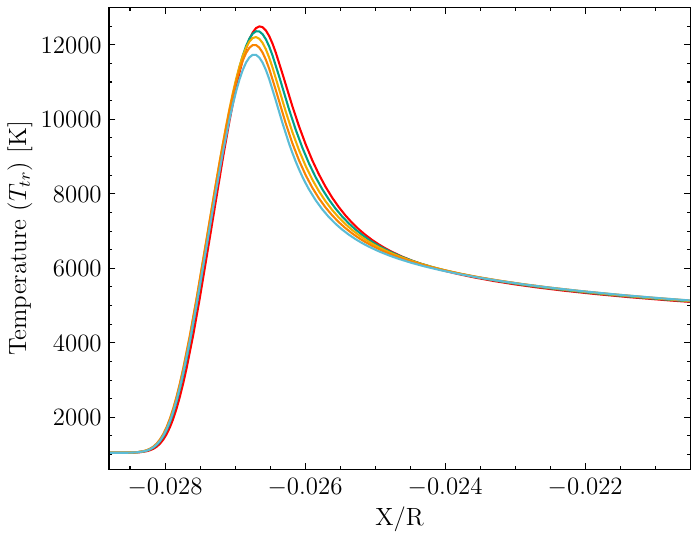}
       \caption{\ce{CO2 + N2} flow.}
       \label{fig:mars_cmp_t_tr_co2_n2}
    \end{subfigure}
    \caption{Mars Pathfinder translational-rotational temperature mode distributions along the stagnation streamline in the non-equilibrium region.}
    \label{fig:mars_cmp_t_tr}
\end{figure}

Figure~\ref{fig:mars_cmp_t_tr} presents the translational-rotational, \( T_{tr} \)\@, temperature mode distributions along the stagnation streamline in the non-equilibrium region for the set of weight factors addressed in this work. Figure~\ref{fig:mars_cmp_t_tr_co2} represents the \( T_{tr} \) temperature mode distributions for the flow composed solely of \ce{CO2} while Fig.~\ref{fig:mars_cmp_t_tr_co2_n2} represents results for the gas mixture of \ce{CO2 + N2}\@. The results shown in this figure indicate that the maximum temperature values decrease with an increase in the \( a \) weight factor value. The \( T_{tr} \) temperature mode distributions obtained in the present work show higher maximum temperature values than the maxima observed in the literature~\cite{moreira_wolf_azevedo_jhmt_2021,lc_scalabrin_phd_2007} for somewhat similar calculations. This result seems to be a direct consequence of the mesh refinement used for the non-equilibrium region in the present calculations.

\begin{figure}[hbt!]
    \centering
    \begin{subfigure}{0.41\linewidth}
       \centering
       \includegraphics[width=\textwidth]{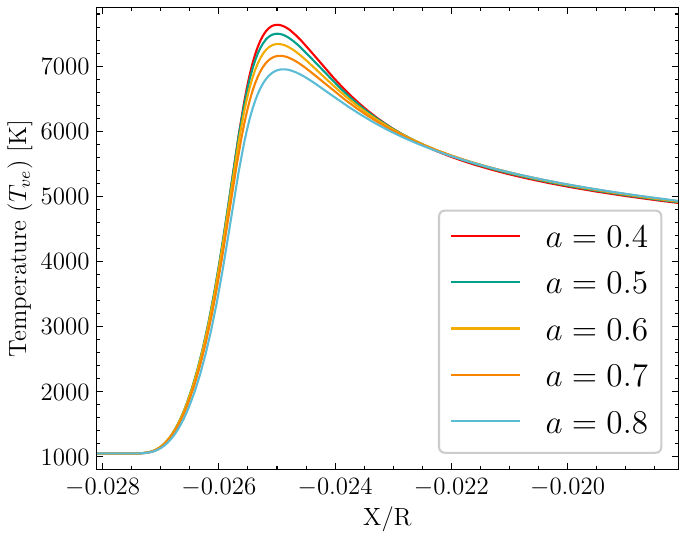}
       \caption{\ce{CO2} flow.}
       \label{fig:mars_cmp_t_ve_co2}
    \end{subfigure}
    \hspace{0.02\textwidth}
    \begin{subfigure}{0.41\linewidth}
       \centering
       \includegraphics[width=\textwidth]{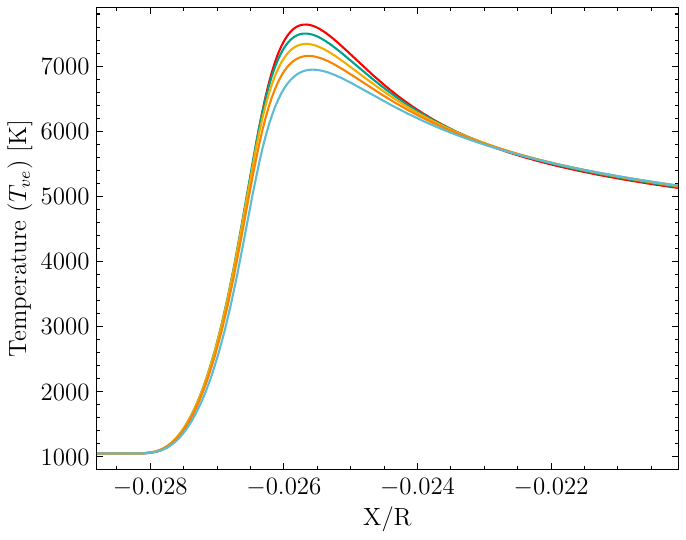}
       \caption{\ce{CO2 + N2} flow.}
       \label{fig:mars_cmp_t_ve_co2_n2}
    \end{subfigure}
    \caption{Mars Pathfinder vibrational-electronic temperature mode distributions along the stagnation streamline in the non-equilibrium region.}
    \label{fig:mars_cmp_t_ve}
\end{figure}

The difference between the maximum \( T_{tr} \) value for \( a = 0.4 \) and \( a = 0.8 \) in Fig.~\ref{fig:mars_cmp_t_tr} is around \SI{1000}{\kelvin} in both cases. The difference between the \ce{CO2} and \ce{CO2 + N2} gas mixtures seems only to be related to the position of the non-equilibrium region, more specifically, the position of the ``ramp up'' of the \( T_{tr} \) temperature mode distributions. Figure~\ref{fig:mars_cmp_t_ve} presents the vibrational-electronic, \( T_{ve} \)\@, temperature mode distributions along the stagnation streamline in the non-equilibrium region for the set of weight factors here used. Figure~\ref{fig:mars_cmp_t_ve_co2} represents the \( T_{ve} \) temperature mode distributions for the flow composed solely of \ce{CO2}, whereas Fig.~\ref{fig:mars_cmp_t_ve_co2_n2} represents results for the gas mixture of \ce{CO2 + N2}\@.

The same behavior observed for the temperature modes in previous results is present in the \( T_{ve} \) temperature mode distributions shown in Fig.~\ref{fig:mars_cmp_t_ve}\@. The vibrational-electronic temperature mode distributions indicate no significant differences when including the \ce{N2} chemical species in the flow gas mixture. This result may relate to the fact that only \SI{5}{\percent} of \ce{N2} exists in the flow gas mixture. Therefore, the energy absorbed by the dissociation reactions of \ce{N2} is not sufficiently large to imply significant changes in the flow behavior.

The simulations regarding the Mars Pathfinder, in both gas mixture conditions, show that the \( T_{tr} \) and \( T_{ve} \) temperature modes do change when varying the weight factors of Park's wo-temperature model. This result is consistent with the behavior observed for the {FIRE II} reentry capsule in previous discussions. However, as stated, this behavior is inconsistent with the results presented in Ref.~\cite{niu_et_al_2018}. The conditions for the Mars Pathfinder simulations in the present work are less severe than those studied in Ref.~\cite{niu_et_al_2018}. This difference in behavior observed, when compared with the literature, indicates that further analyses regarding Park's control temperature weight factors are necessary.

\subsubsection{Stagnation Point Convective Heat Flux}

\begin{figure}[hbt!]
    \centering
    \begin{subfigure}{0.41\linewidth}
       \centering
       \includegraphics[width=0.8\textwidth]{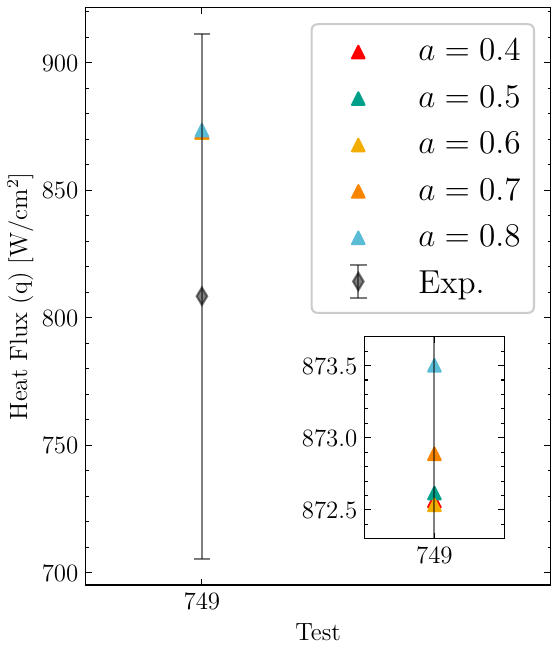}
       \caption{\ce{CO2} flow.}
       \label{fig:mars_cmp_q_co2}
    \end{subfigure}
    \hspace{0.02\textwidth}
    \begin{subfigure}{0.41\linewidth}
       \centering
       \includegraphics[width=0.8\textwidth]{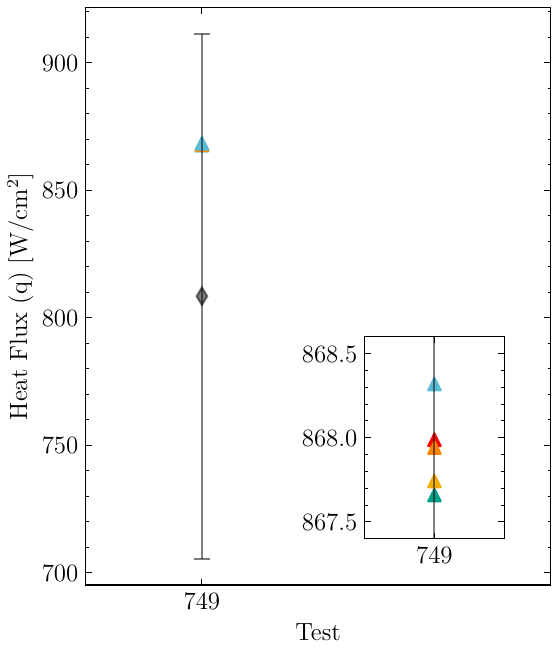}
       \caption{\ce{CO2 + N2} flow.}
       \label{fig:mars_cmp_q_co2_n2}
    \end{subfigure}
    \caption{Mars Pathfinder stagnation point convective heat flux.}
    \label{fig:mars_cmp_q}
\end{figure}

Figure~\ref{fig:mars_cmp_q} shows the stagnation point convective heat flux for the different sets of weight factors proposed in the present work. Figure~\ref{fig:mars_cmp_q_co2} presents the results for the flow composed solely of \ce{CO2}\@, and Fig.~\ref{fig:mars_cmp_q_co2_n2} has the results for the flow gas mixture of \ce{CO2 + N2}\@. The experimental data used as reference are from Ref.~\cite{hollis_1996}, represented by the black diamond-shaped symbol. The margin of error, represented by the black bars, is approximately \SI{12.7}{\percent} for the total stagnation heat flux. As stated, this experimental data is only for carbon dioxide flows.

The results obtained in this work for both flow conditions are within the margin of error of the experimental data, as shown in Fig.~\ref{fig:mars_cmp_q}\@. The differences between the results for each set of weight factors are less than \SI{0.15}{\percent} of the average predicted value for both gas mixtures. The present simulations overpredict the stagnation point convective heat flux when compared to the experimental data. The flow with the \ce{CO2 + N2} gas mixture yields lower stagnation point convective heat flux values. This difference is due to dissociation reactions of the \ce{N2}, which absorb energy available in the flowfield. Note that the present analysis does not include the radiative phenomena. Therefore, the impact of Park's control temperature weight factors in predicting the stagnation point convective heat flux is significantly small and can be considered negligible.

\subsubsection{Chemical Species Mass Fraction}

\begin{figure}[hbt!]
    \centering
    \begin{subfigure}{0.41\linewidth}
       \centering
       \includegraphics[width=\textwidth]{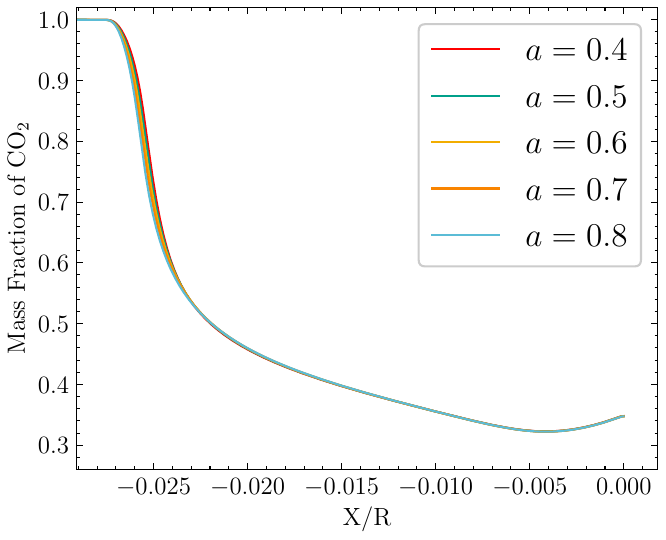}
       \caption{\ce{CO2} flow.}
       \label{fig:mars_cmp_chemcial_species_CO2_co2}
    \end{subfigure}
    \hspace{0.02\textwidth}
    \begin{subfigure}{0.41\linewidth}
       \centering
       \includegraphics[width=\textwidth]{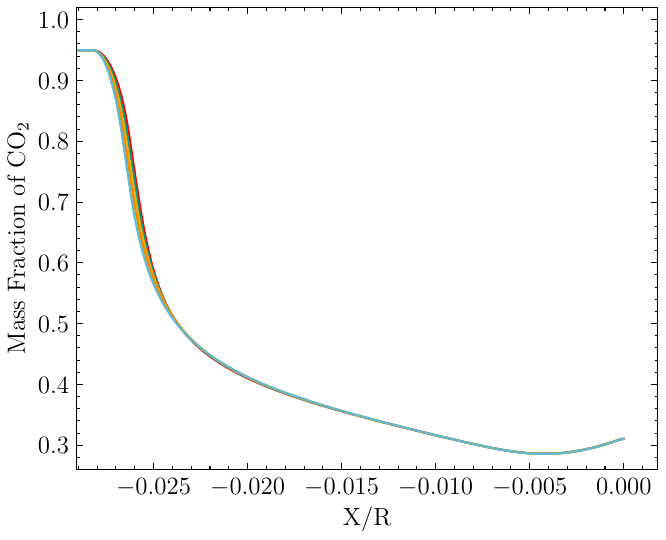}
       \caption{\ce{CO2 + N2} flow.}
       \label{fig:mars_cmp_chemcial_species_CO2_co2_n2}
    \end{subfigure}
    \caption{Mars Pathfinder \ce{CO2} mass fraction distributions along the stagnation streamline in the non-equilibrium region.}
    \label{fig:mars_cmp_chemcial_species_CO2}
\end{figure}

The present study considers an 8-species chemical model to simulate Mars' atmosphere. Moreover, this work proposed an analysis of two different flow gas mixture compositions. Therefore, presenting the mass fraction distributions for all chemical species considered in this study would be excessive detail for the present conference paper. Thus, the results presented in this section are related to the chemical species considered strong radiators by the literature~\cite{boyd_schwartzentruber_2017,mf_modest_2013,moreira_levin_thirani_wolf_azevedo_scitech_2024,priyadarshini_et_al_aiaa_jthtc_2018}\@. The chemical species chosen are \ce{CO2}\@, \ce{CO}\@, \ce{N}\@, and \ce{NO}\@. The mass fraction distributions for \ce{N} and \ce{NO} are only present in the gas mixture composition of \ce{CO2 + N2}\@. Figure~\ref{fig:mars_cmp_chemcial_species_CO2} shows the \ce{CO2} mass fraction distributions along the stagnation streamline in the non-equilibrium region for the set of weight factors \( a = 0.4 \)\@, \( 0.5 \)\@, \( 0.7 \)\@, \( 0.8 \)\@, and \( 0.8 \)\@. The left graph in Fig.~\ref{fig:mars_cmp_chemcial_species_CO2} refers to the gas mixture composition of \ce{CO2}\@, whereas the right one refers to the \ce{CO2 + N2} gas mixture composition.

\begin{figure}[hbt!]
    \centering
    \begin{subfigure}{0.41\linewidth}
       \centering
       \includegraphics[width=\textwidth]{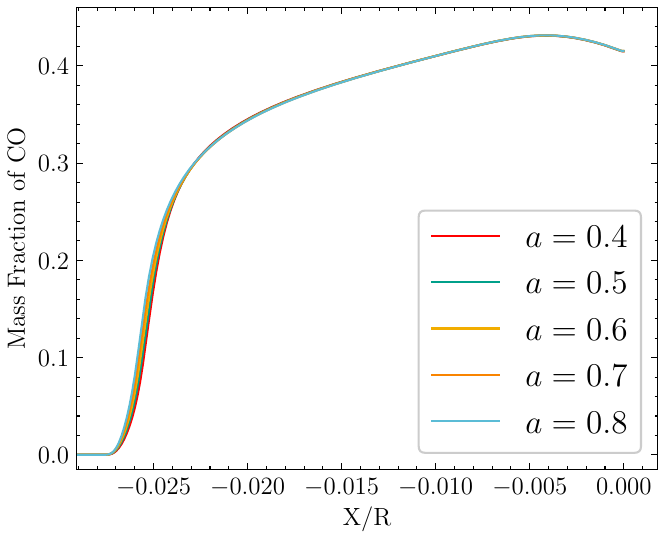}
       \caption{\ce{CO2} flow.}
       \label{fig:mars_cmp_chemcial_species_CO_co2}
    \end{subfigure}
    \hspace{0.02\textwidth}
    \begin{subfigure}{0.41\linewidth}
       \centering
       \includegraphics[width=\textwidth]{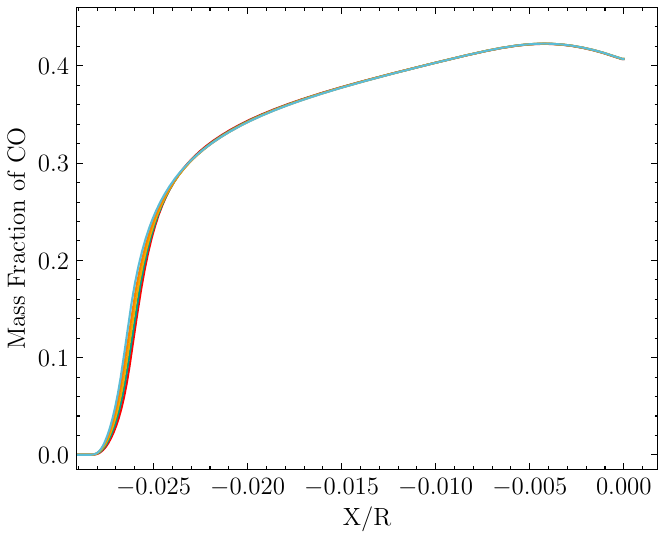}
       \caption{\ce{CO2 + N2} flow.}
       \label{fig:mars_cmp_chemcial_species_CO_co2_n2}
    \end{subfigure}
    \caption{Mars Pathfinder \ce{CO} mass fraction distributions along the stagnation streamline in the non-equilibrium region.}
    \label{fig:mars_cmp_chemcial_species_CO}
\end{figure}

The \ce{CO2} mass fraction distributions in Fig.~\ref{fig:mars_cmp_chemcial_species_CO2} are very similar between the two different gas mixture compositions. This result indicates that the addition of \ce{N2} does not impact the behavior of the \ce{CO2} mass fraction distributions. Figure~\ref{fig:mars_cmp_chemcial_species_CO2} shows that, in both gas mixture compositions, the increase of the \( a \) weight factor accelerates the consumption of \ce{CO2} in the non-equilibrium region. This result follows the expected behavior of endothermic reactions, where control temperature distributions, \( T_{c} \)\@, of higher magnitudes favor dissociation reactions, which are endothermic chemical reactions. Moreover, these results agree with the observed behavior in the simulation for the {FIRE II} reentry capsule. Inside the shock layer, the mass fraction distributions tend to ``converge'' to a unique solution with different sets of weight factors. These results indicate that even with the flow gas mixture changing along the stagnation streamline, the flow has sufficient relaxation time to achieve the equilibrium state. Estimating the \ce{CO2} mass fraction distributions is of interest because it influences radiative phenomena. Figure~\ref{fig:mars_cmp_chemcial_species_CO} shows the \ce{CO2} mass fraction distributions along the stagnation streamline in the non-equilibrium region for the set of weight factors \( a = 0.4 \)\@, \( 0.5 \)\@, \( 0.7 \)\@, \( 0.8 \)\@, and \( 0.8 \)\@. the left graph in Fig.~\ref{fig:mars_cmp_chemcial_species_CO} refers to the gas mixture composition of \ce{CO2} and the right graph refers to the \ce{CO2 + N2} gas mixture composition.

\begin{figure}[hbt!]
    \centering
    \begin{subfigure}{0.41\linewidth}
       \centering
       \includegraphics[width=\textwidth]{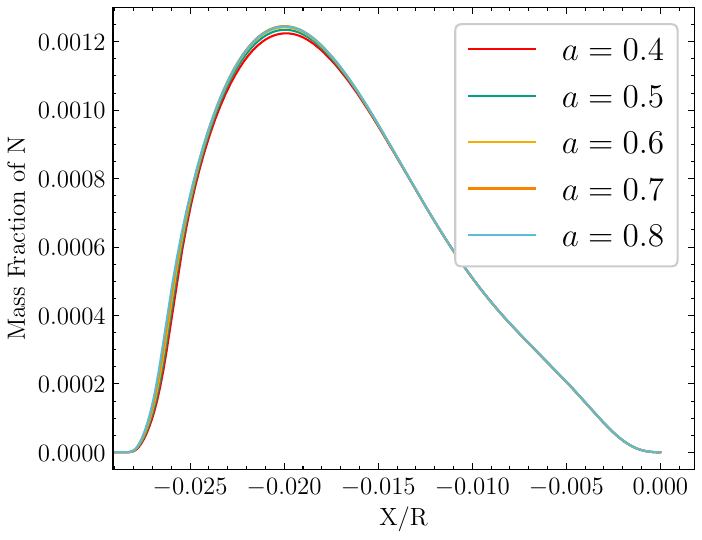}
       \caption{\ce{N} mass fraction distributions.}
       \label{fig:mars_cmp_chemcial_species_N}
    \end{subfigure}
    \hspace{0.02\textwidth}
    \begin{subfigure}{0.41\linewidth}
       \centering
       \includegraphics[width=\textwidth]{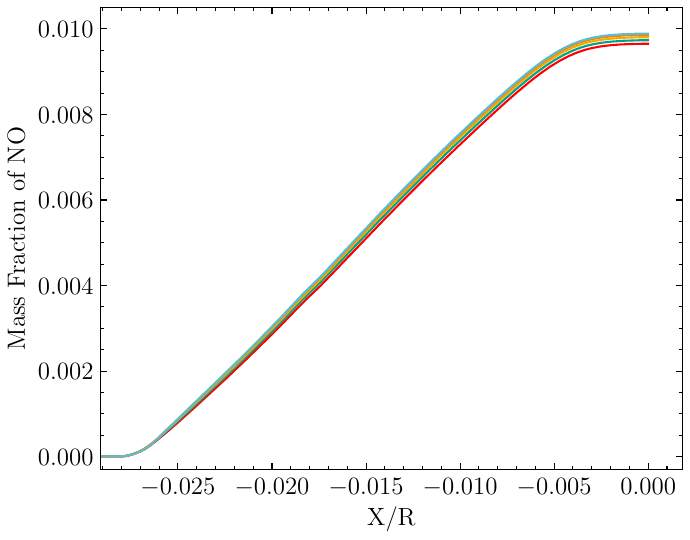}
       \caption{\ce{NO} mass fraction distributions.}
       \label{fig:mars_cmp_chemcial_species_NO}
    \end{subfigure}
    \caption{Mars Pathfinder \ce{N} and \ce{NO} mass fraction distributions along the stagnation streamline in the non-equilibrium region for the \ce{CO2 + N2} freestream gas mixture composition.}
    \label{fig:mars_cmp_chemcial_species_co2_n2}
\end{figure}

The \ce{CO} mass fraction distributions, Fig.~\ref{fig:mars_cmp_chemcial_species_CO}\@, are, in a metaphorical way, a specular reflection of the \ce{CO2} mass fraction distributions, Fig.~\ref{fig:mars_cmp_chemcial_species_CO2}\@. This behavior is expected because the \ce{CO2} dissociates into \ce{CO + O} by the chemical reaction
\begin{equation}
    \ce{CO2 + M <=> CO + O + M}
    \qquad \text{,}
\end{equation}
where \( \text{M} \) represents the reaction partner that can be any other chemical species considered in the chemical model. The increase in the \( a \) weight factor value accelerates the production of \ce{CO}\@. The changes in the weight factor values do not impact the shock layer, as expected for a fluid in the equilibrium state. Note that the differences caused by the different values of the weight factor in the mass fraction distributions in the non-equilibrium region are small in both Figs.~\ref{fig:mars_cmp_chemcial_species_CO2} and \ref{fig:mars_cmp_chemcial_species_CO}\@. However, the gap between the mass fraction distributions within the non-equilibrium region is consistent, meaning that there is no unique solution. Figure~\ref{fig:mars_cmp_chemcial_species_co2_n2} shows the \ce{N} and \ce{NO} mass fraction distributions along the stagnation streamline in the non-equilibrium region for the set of weight factors \( a = 0.4 \)\@, \( 0.5 \)\@, \( 0.7 \)\@, \( 0.8 \)\@, and \( 0.8 \)\@. The left graph, Fig.~\ref{fig:mars_cmp_chemcial_species_N}\@, refers to the mass fraction distributions of atomic nitrogen, and the right one, Fig.~\ref{fig:mars_cmp_chemcial_species_NO}\@, refers to the \ce{NO} mass fraction distributions.

Figure~\ref{fig:mars_cmp_chemcial_species_N} shows the production and consumption of \ce{N} in the non-equilibrium region and shock layer. The increase of the \( a \) weight factor value accelerates the production of \ce{N} by favoring the dissociation of \ce{N2}\@. The maximum value of \ce{N} increases with the increase of the \( a \) value. However, the differences among the maximum values of \ce{N} mass fraction for \( a = 0.6 \)\@, \( 0.7 \)\@, and \( 0.8 \) are quite small. Moreover, the gap between the mass fraction distributions of different sets of weight factors is also small. The mass fraction distributions in the shock layer ``converge'' to a unique solution, indicating the equilibrium state. Figure~\ref{fig:mars_cmp_chemcial_species_NO} shows that the increase in the \( a \) weight factor value yields higher mass fraction values. The differences between the mass fraction distributions are consistent, indicating that the different sets of weight factors do not ``converge'' to a unique solution. The \ce{NO} mass fraction increases along the shock layer. The results observed for the \ce{N} and \ce{NO} chemical species are consistent with the theoretical formulation adopted by the numerical tool.

\section{CONCLUDING REMARKS}

The present study investigated hypersonic entry/reentry flows in thermodynamic non-equilibrium conditions. The Navier-Stokes equations with source terms accounting for chemical reactions and non-equilibrium phenomena are solved numerically. The present work simulated flow conditions for Earth reentry and Mars entry hypersonic flows in thermodynamic non-equilibrium. An 11-species chemical model simulates Earth's atmosphere, and an 8-species chemical model simulates Mars' atmosphere. The present work used a numerical tool with Park's two-temperature model implemented to account for the thermal non-equilibrium. The weight factors used to calculate the control temperature, \( T_{c} \)\@, in Park's two-temperature model, ``controls'' the energy transfer between dissociation and ionization reactions. The present work analyzed the influence of the choice of the set of weight factors of Park's two-temperature model on the flow behavior of Mars and Earth's hypersonic entry/reentry flows. The results presented in this study are in terms of the Mach number, translational-rotational and vibrational-electronic temperature modes, and chemical species mass fraction distributions along the stagnation streamline. Moreover, the present study also presents results for the stagnation point convective heat flux on the vehicle body. The present work results agree with experimental and numerical data in the literature.

The present study aims to broaden the understanding regarding the influence of Park's two-temperature model weight factors on the flow behavior. A better understanding of Park's weight factors allows for better use of Park's model. The results observed in the present study show the impact of the choice of the set of weight factors of Park's two-temperature model on the flow behavior. The changes caused by the weight factors occur in regions of the domain with high-intensity non-equilibrium behavior, such as the region around the shock wave. The increase of the \( a \) weight factor tends to move the shock wave away from the vehicle body in all cases tested. However, the changes are small compared to the position of the shock wave in relation to the vehicle body. The present work shows that the weight factors greatly influence the \( T_{tr} \) and \( T_{ve} \) temperature mode distributions for both Earth and Mars hypersonic flows. There is a slight change in the position of the temperature mode distributions, where the temperature mode distributions tend to move away from the vehicle body. Moreover, the increase of the \( a \) weight factor causes a decrease in the maximum temperature values observed. The magnitude of the temperature mode maximum values can vary by thousands of Kelvin. The changes in the temperature profiles are closely related to chemical reactions.

The control temperature directly affects the mass fraction distributions of the chemical species considered in the flow. As presented in the theoretical formulation, Park's control temperature is the reference to calculate the chemical reaction rates and the equilibrium constant. The increase of the \( a \) weight factor value increases the rate of consumption or production of the chemical species that participate in dissociation reactions. The present work found that the different sets of weight factors affect each chemical species' mass fraction distributions differently based on the chemical reactions involved with the chemical species. The impact of the choice of the set of weight factors is more visible and significant in the hypersonic flow regarding the {FIRE II} reentry capsule. In the simulations performed for the {FIRE II} capsule, the present work found that the weight factors significantly influence the \ce{NO} mass fraction distributions. This behavior is mainly by the dissociation reaction of \ce{NO}\@, favored by higher values of the \( a \) weight factor, and the production of \ce{NO} limited by chemical reactions not influenced by the changes in the weight factor values.

The present work also evaluated the stagnation point convective heat flux for the {FIRE II} and Mars Pathfinder flow conditions. The results show good agreement with the experimental data used as a reference. The changes in the weight factors seem to follow a consistent behavior for the simulations performed for the {FIRE II} reentry capsule at the conditions of \( H = \SI{71.02}{\km} \) and \( \SI{41.60}{\km} \)\@, while the same is not valid for \( H = \SI{48.37}{\km} \)\@. For the Mars Pathfinder simulations, the results regarding the stagnation point convective heat flux do not seem to follow a pattern. Therefore, the variations in the temperature distributions along the stagnation line and closer to the stagnation point may arise from numerical fluctuations for the Mars Pathfinder simulations. For the {FIRE II} cases, the authors are still investigating the patterns observed in the stagnation point convective heat flux results for the {FIRE II} simulations. In both {FIRE II} and Mars Pathfinder simulations, the differences between the stagnation point convective heat flux results are less than \SI{0.1}{\percent} of the average value of the results for each case. Thus, the impact of the weight factor values is not of the same significance when compared with the changes observed in the non-equilibrium region.

Reference~\cite{c_park_2010} indicate that there is no advantage of using \( a = 0.7 \) over \( a = 0.5 \)\@. Moreover, Ref.~\cite{niu_et_al_2018} found that the comparison between \( a = 0.7 \) and \( a = 0.5 \) produces visible differences in the vibrational-electronic temperature mode while no changes are observed for the translational-rotational temperature mode distributions. The present work results show significant changes in the property distributions within the non-equilibrium region. Therefore, the choice of the set of weight factors may be important, depending on the flow conditions and research goals. However, considering the data collected in the present work, the authors cannot recommend any specific weight factor values for the correct solution of the non-equilibrium region. It seems that the sets of the weight factors, typically suggested by the literature, are indeed adequate for some simulations. However, as shown in the results for the {FIRE II}\@, there are some cases where the weight factor values significantly impact the property distributions. With this in mind, more data would be necessary to specify the correct sets of weight factors. Moreover, the flow conditions must be accounted for when choosing the weight factor values.

\section{ACKNOWLEDGEMENTS}

The authors gratefully acknowledge the support for the present research provided by Fundação de Amparo à Pesquisa do Estado de São Paulo, FAPESP, under the Research Grants No.\  2013/07375-0\@, 2021/02705-8\@, 2022/07604-8\@, and 2024/07590-2\@. The availability of computational resources from the National Laboratory for Scientific Computing, LNCC/MCTI, under the HFWBTF Project at the SDumont supercomputer, and from the Center for Mathematical Sciences Applied to Industry, CEPID-CeMEAI, also funded by FAPESP under the Research Grant No. 2013/07375-0\@, are also gratefully acknowledged. Partial support for the present research was also provided by Conselho Nacional de Desenvolvimento Científico e Tecnológico, CNPq, under the Research Grant No.\ 315411/2023-6\@. This study was financed in part by Coordenação de Aperfeiçoamento de Pessoal de Nível Superior - Brasil (CAPES) - Finance Code 001\@.

\section{REFERENCES} 

\renewcommand{\refname}{}
\bibliography{bibfile.bib}

\end{document}